\documentclass[a4paper,11pt]{article}
\usepackage{jheppub} % for details on the use of the package, please
\usepackage{caption}
\usepackage{subcaption}
\usepackage{graphicx,array}
\usepackage{color}
\usepackage{setspace}
\usepackage{amstext}
\usepackage{amssymb}
\usepackage{slashed}
\usepackage{makecell}
\usepackage{multirow}
\usepackage{wasysym}
\usepackage{physics}
\usepackage[T1]{fontenc} % if needed
\usepackage{changepage}
\usepackage{arydshln}
\usepackage{mathrsfs} %Ralph Smith’s Formal Script Font (rsfs):Use the “mathrsfs” package. 
\RequirePackage{luatex85}
\usepackage{tikz-feynman}
\tikzset{graviton/.style={decorate, decoration={snake, amplitude=.4mm, segment length=1.5mm, pre length=.5mm, post length=.5mm}, double}}
\usepackage{scalerel} %To make the size of super/subscript smaller      

\definecolor{orange}{rgb}{1,0.5,0}
\definecolor{brown}{rgb}{0.59, 0.29, 0.0}

\definecolor{note_fontcolor}{rgb}{0.80078125, 0.80078125, 0.80078125}

\definecolor{darkgreen}{rgb}{0,0.5,0}

\def\beq{\begin{equation}}
\def\eeq{\end{equation}}
\def\bea{\begin{eqnarray}}
\def\eea{\end{eqnarray}}

\def\beq{\begin{equation}}
\def\eeq#1{\label{#1}\end{equation}}
\def\eeqn{\end{equation}}
\def\beqa{\begin{eqnarray}}
\def\eeqa#1{\label{#1}\end{eqnarray}}
\def\eeqan{\end{eqnarray}}
\def\CR{\nonumber \\ }
\def\leqn#1{(\ref{#1})}

\newcommand{\centeron}[2]{{\setbox0=\hbox{#1}\setbox1=\hbox{#2}\ifdim
		\wd1>\wd0\kern.5\wd1\kern-.5\wd0\fi \copy0
		\kern-.5\wd0\kern-.5\wd1\copy1\ifdim\wd0>\wd1
		\kern.5\wd0\kern-.5\wd1\fi}}
\newcommand{\ltap}{\>\centeron{\raise.35ex\hbox{$<$}}
	{\lower.65ex\hbox{$\sim$}}\>}
\newcommand{\gtap}{\>\centeron{\raise.35ex\hbox{$>$}}
	{\lower.65ex\hbox{$\sim$}}\>}
\newcommand{\gsim}{\mathrel{\gtap}}

\def\yi{y_{\rm IR}}

\makeatletter
\newcommand*{\Relbarfill@}{\arrowfill@\Relbar\Relbar\Relbar}
\newcommand*{\xeq}[2][]{\ext@arrow 0055\Relbarfill@{#1}{#2}}
\makeatother

\usepackage{cleveref}

\title{\boldmath Collider Signatures of Near-Continuum Dark Matter}

\author[a]{Steven Ferrante,}
\author[b]{Seung J. Lee,}
\author[a]{and Maxim Perelstein}

\affiliation[a]{Department of Physics, LEPP, Cornell University, Ithaca, NY 14853, USA}
\affiliation[b]{Department of Physics, Korea University, Seoul, 136-713, Korea}

\abstract{In this paper we study a near-continuum dark matter model, in which dark sector consists of a tower of closely spaced states with weak-scale masses. We construct a five-dimensional model which naturally realizes this spectrum. The dark matter is described by a bulk field, which interacts with the brane-localized Standard Model sector via a $Z$ portal. We then study collider signatures of this model. Near-continuum dark matter states produced in a collider undergo cascade decays, resulting in events with high multiplicity of jets and leptons, large missing energy, and displaced vertices. A custom-built Monte Carlo tool described in this paper allows for detailed simulation of the signal events. We present results of such simulations for the case of electron-positron collisions.}

\begin{document}
\maketitle
\flushbottom

\section{Introduction}
\label{sec:Intro}

While existence of dark matter (DM) is well-established, its microscopic nature remains a mystery, motivating theoretical investigation of a broad range of possible candidate models. Models that predict novel observational or experimental signatures of dark matter are of particular interest, as they may motivate new experiments or search strategies. Recently, a new {\it Continuum Dark Matter} framework was proposed~\cite{Csaki:2021gfm,Csaki:2021xpy,Csaki:2022lnq}, in which dark matter consists of gapped continuum states, rather than ordinary particles. While no gapped continuum states have yet been seen experimentally in high-energy physics, they do appear in a variety of quantum field theories (QFTs)~\cite{McCoy:1978ta, McCoy:1978ix,Wu:1977hi,Luther:1976mt,Cabrer:2009we}, as well as in condensed matter systems~\cite{Fradkin:1991nr, sachdev2007quantum}.\footnote{For other applications of gapped continuum in Beyond the Standard Model phenomenology, see Refs.~\cite{Falkowski:2008fz,Stancato:2008mp,Falkowski:2008yr,Falkowski:2009uy,Bellazzini:2015cgj,Katz:2015zba,Csaki:2018kxb,Megias:2019vdb,Megias:2021mgj,Fichet:2022ixi,Chaffey:2021tmj,Aoki:2023tjm}. And also see~\cite{Kumar:2018jxz,Gabadadze:2021dnk,Eroncel:2023uqf} for how gapped continuum arises in the context of certain inflationary cosmology scenarios.} Explicit models constructed in~\cite{Csaki:2021gfm,Csaki:2021xpy,Csaki:2022lnq} are based on local, unitary five-dimensional (5D) QFTs, which contain fields with gapped-continuum spectra. It was shown that excitations of these fields can play the role of cold dark matter, while satisfying all known observational and experimental constraints. The Continuum DM framework predicts new phenomena, that do not appear in ordinary particle DM models: continuous decay of DM states throughout the cosmological history; strong suppression of direct-detection cross sections due to continuum kinematics; and cascade decays of DM produced at colliders. The goal of this paper is to investigate the collider phenomenology of this class of models in more detail.   

The 5D models studied in Ref.~\cite{Csaki:2021gfm,Csaki:2021xpy,Csaki:2022lnq}, based on the soft-wall geometry originally introduced in Ref.~\cite{Cabrer:2009we}, contain a naked singularity at a finite distance from the location of the 4D brane on which the Standard Model (SM) fields are confined.\footnote{Note that this singularity is classified as a ``good" naked singularity~\cite{Gubser:2000nd}} Near the singularity, the effective field theory description of the space-time and fields propagating on it breaks down, and needs to be supplanted with a more fundamental description incorporating quantum gravity, such as string theory, which is expected to smooth out (or resolve) the singularity.\footnote{In the context of string theory, the presence of a gapped continuum arises in scenarios involving a significant quantity of $D$3 branes distributed on a disk. which is dual to $N = 4$ supersymmetry (SUSY) broken down to $N = 2$ SUSY, achieved by introducing masses for two chiral adjoints. A considerable body of literature has explored this topic, for example see~\cite{Gubser:2000nd,Freedman:1999gk,Kraus:1998hv}.}
In this paper, we model the effects of this extra physics by introducing an infrared (IR)-regulator end-of-space brane, which cuts off the singular region of space-time. In the presence of the IR-regulator brane, 5D fields appear as discrete Kaluza-Klein (KK) towers from the 4D point of view.  However, if the regulator brane is placed close to the singularity, the mass splitting between the neighboring KK modes is much smaller than other physical scales, such as the gap scale. Models with such spectra, which we term {\it near-continuum}, share many of the features of the continuum models considered in~\cite{Csaki:2021gfm,Csaki:2021xpy,Csaki:2022lnq}. In particular, the mass splittings between modes may be sufficiently small that their collider phenomenology is well modeled by a continuous spectrum. This is the approach that will be taken in this work. Since general-purpose Monte Carlo (MC) tools used in collider phenomenology, such as {\tt MadGraph}~\cite{Alwall:2011uj}, do not include continuous spectra\footnote{Note also that using the existing general-purpose tools to model discrete spectra with a large number of states is computationally impractical; see Sec.~\ref{subsec:NCProduction}.}, we constructed a custom-made MC tools to study the collider signatures of the model at hand.  We then study the phenomenology of a near-continuum DM model in $e^+e^-$ collisions at $\sqrt{s}=500$~GeV. This allows us to describe the characteristic signatures of the model in a simple setting, and demonstrates the use of our MC tool. Near-continuum DM signatures at hadron colliders, such as the LHC, are expected to be similar, and will be studied in future work. 

Theoretical and phenomenological aspects of multi-component dark matter were investigated in a broad framework termed Dynamical Dark Matter (DDM) by Dienes, Thomas, and their collaborators~\cite{Dienes:2012yz,Dienes:2012cf,Dienes:2013xya,Dienes:2014via,Dienes:2014bka,Boddy:2016fds,Curtin:2018ees,Dienes:2019krh,Dienes:2021cxr,Dienes:2022zbh}. The relation between the Continuum DM framework and DDM was discussed in Ref.~\cite{Csaki:2021gfm}. In terms of collider signatures, the two frameworks bear many similarities, such as the appearance of high-multiplicity final states~\cite{Dienes:2019krh} and multiple displaced vertices~\cite{Dienes:2021cxr} from cascade decays. In the near-continuum DM model, multi-component DM emerges naturally from a simple and well-motivated 5D model, providing a highly-predictive setup in which properties of the signal can be predicted in detail. Making such predictions is further facilitated by the custom-built Monte Carlo simulation tool constructed in this paper.  
In the future, it may be interesting to compare the predictions of the two frameworks and understand how they may be distinguished should a collider signal with the characteristic features be observed. 

The rest of the paper is organized as follows. Section~\ref{sec:model} reviews the soft-wall geometry and the $Z$-portal weak-interacting continuum model of dark matter. The IR-regulator brane is then introduced, and the resulting near-continuum DM spectrum is discussed. Section~\ref{sec:NCProdDecay} contains the discussion of production of near-continuum DM states in $e^+e^-$ collisions, and their subsequent cascade decay. Results of a Monte Carlo study of this process are presented in Section~\ref{sec:VisiblePheno}. Finally, Section~\ref{sec:conc} contains outlook for future studies and concluding remarks. The custom-built Monte Carlo tool used to model production of decay of states with continuous spectra is described in the Appendix.     

\section{Near-Continuum Dark Matter from a 5D Soft-Wall Model}	
\label{sec:model}

\subsection{5D Model of Continuum DM}

Here we briefly review the 5D space constructed in Ref.~\cite{Cabrer:2009we}, and the model of dark matter, based on this geometry, presented in Refs.~\cite{Csaki:2021gfm,Csaki:2021xpy}.

Consider a 5D space parametrized by $(x^\mu, y)$, and assume Poincare invariance in the four dimensions $x^\mu$, so that the metric has the form $ds^2=e^{2A(y)}\eta_{\mu\nu}dx^\mu dx^\nu - dy^2$. A scalar field $\varphi$, minimally coupled to gravity, can propagate on this space. A single 4D brane is located at $y=0$, and we impose the orbifold $Z_2$ symmetry $y\to -y$ under which both the metric and the scalar field are even. The action is given by 
\beq
S = \int d^5 x \sqrt{-g} \left(M_5^3 R - 3(\partial \varphi)^2 - V[\varphi]\right) - \int d^4x \sqrt{-g_{\rm ind}} \lambda[\varphi(y=0)].
\eeq{5Daction}
Here $M_5$ is the 5D Planck scale, while $V[\varphi]$ and $\lambda[\varphi]$ are the bulk and brane contributions, respectively, to the scalar potential. We set $M_5=1$ in the remainder of this section. An exact solution for the coupled scalar-metric classical equations of motion can be obtained in closed form provided that the scalar potentials $V$ and $\lambda$ can be written in terms of a ``superpotential" $W[\varphi]$ as
\beqa
V[\varphi] = 3 \left( \frac{\partial W}{\partial \varphi}\right)^2\,-\,12 W^2;~~~~~~\lambda[\varphi]=6 W[\varphi].
\eeqa{ass}
Following Ref.~\cite{Cabrer:2009we}, we choose a simple superpotential
\beq
W = k \, \left(1+e^\varphi\right)\,,
\eeq{W}
which yields a classical solution
\beq
A(y)\,=\, ky-\log\left( 1-\frac{y}{y_s}\right);~~~~~~\varphi(y) \,=\, -\log k(y_s-y).
\eeq{BG}     
Here $y_s$ is an integration constant. This metric has a curvature singularity at $y=y_s$, which is interpreted as indicating that the spacetime ends at $y_s$. We will generally consider the regime
\beq
x_s \equiv ky_s \gg 1.
\eeq{kys}
In this case, the geometry is approximately that of an AdS slice (as in Randall-Sundrum models) for $y\ll y_s$, with significant deviations kicking in as $y$ approaches $y_s$. We will refer to this metric as {\it soft-wall geometry}; it is represented schematically in Fig.~\ref{fig:NearContinuumSetup}. 

Small fluctuations in the metric-scalar system around the ground state of Eq.~\leqn{BG} are identified as the graviton (tensor) and radion (scalar) fields. The graviton spectrum contains a zero-mode, so that at large distances, 4D Newtonian gravity is reproduced. The 4D Planck scale is given by
\beq
M_4^2 = \frac{M_5^3}{2k}\,\left(1-\frac{1}{x_s}+\frac{1}{2x_s^2}-\frac{1}{2}e^{-2x_s}\right)\,.
\eeq{MPl4D}     
Kaluza-Klein (KK) excitations of the graviton form a gapped continuum, with the minimal (gap) 4D mass given by
\beq
\mu_0 = \frac{3}{2}\,\frac{e^{-ky_s}}{y_s}.
\eeq{gap}
The radion excitation does not have a zero mode, but has a gapped continuum starting at the same $\mu_0$. We will be interested in the parameter region where $M_5$, $k$ and $y_s$ are all within a few orders of magnitude of the 4D Planck scale, but $\mu_0$ is at the weak scale, $\sim 100$~GeV. This is natural given a mild ${\cal{O}}(10)$ hierarchy between $y_s$ and $1/k$. 

The Standard Model (SM) fields can be incorporated in several ways. For example, they can be localized on the $y=0$ brane, or be interpreted as zero-modes of bulk fields. Here, we choose to introduce an additional 4D brane at $y=y_b < y_s$, with $k y_b\sim {\cal O}(10)$. SM fields are localized on this brane; see Fig.~\ref{fig:NearContinuumSetup}. This setup has the advantage of preserving a Randall-Sundrum (RS)-like solution to the gauge hierarchy problem: the mass scales on the SM brane are exponentially suppressed by the warp factor~\cite{Randall:1999ee}. We assume that the tension of the SM brane is small enough to not significantly perturb the metric.        

In Refs.~\cite{Csaki:2021gfm,Csaki:2021xpy,Csaki:2022lnq}, it was suggested that dark matter (DM) can be described by a scalar field $\Phi$ propagating on the classical geometry of Eq.~\leqn{BG}. A discrete $Z_2$ symmetry, under which $\Phi\to -\Phi$, is imposed to ensure DM stability on cosmological time scales. The bulk action is 
\beq
S_{\rm DM} = \int d^5 x \sqrt{-g} \left(g^{MN}\partial_M \Phi \partial_N \Phi  - m^2 \Phi^2\right)\,,
\eeq{L_DM}
and the scalar field is assumed to propagate in the region $y_b \leq y < y_s$, {\it i.e.} between the SM brane and the singularity. From the 4D perspective, the field $\Phi$ appears as a gapped continuum, with the same gap scale $\mu_0$ as in Eq.~\leqn{gap}, independent of the bulk mass $m$. As an example of an explicit and viable model of DM, 
Refs.~\cite{Csaki:2021gfm,Csaki:2021xpy} considered a ``$Z$-portal" interaction of the $\Phi$ field with the SM:
\beq
S_{\rm DM-SM} = \int d^4 x \sqrt{-g_{\rm ind}} \left( g^{MN} D_M\chi^\dagger D_N\chi - m_\chi^2|\chi|^2-\hat{\lambda} k^{1/2} \Phi(y_b) \chi H +~{\rm h.c.}\right),
\eeq{L_DMSM}
where $H$ is the SM Higgs doublet, and $\chi$ is an additional weak-doublet field localized on the SM brane. The $\chi$ field is odd under the DM $Z_2$ symmetry, while all SM fields are even.  Electroweak symmetry breaking introduces a mass mixing between $\chi^0$ and $\Phi$, and if $\chi^0$ is integrated out, an effective coupling of $\Phi(y_b)$ to the SM $W$ and $Z$ bosons is produced. The coupling has the usual structure required by the 4D gauge invariance, 
with the coupling strength to the $Z$ given by
\beq
g_{\rm eff}=g_Z \sin^2\alpha
\eeq{geff}
where $g_Z=\sqrt{g^2+g^{\prime 2}}$ is the SM $Z$ coupling constant, and the mixing angle is in the range $\sin^2\alpha\sim 0.1-0.01$ for phenomenologically viable DM models. For further details, see Ref.~\cite{Csaki:2021gfm}.  

\subsection{Regulator Brane and Near-Continuum DM}

\begin{figure}
\centering
\includegraphics[width=4in]{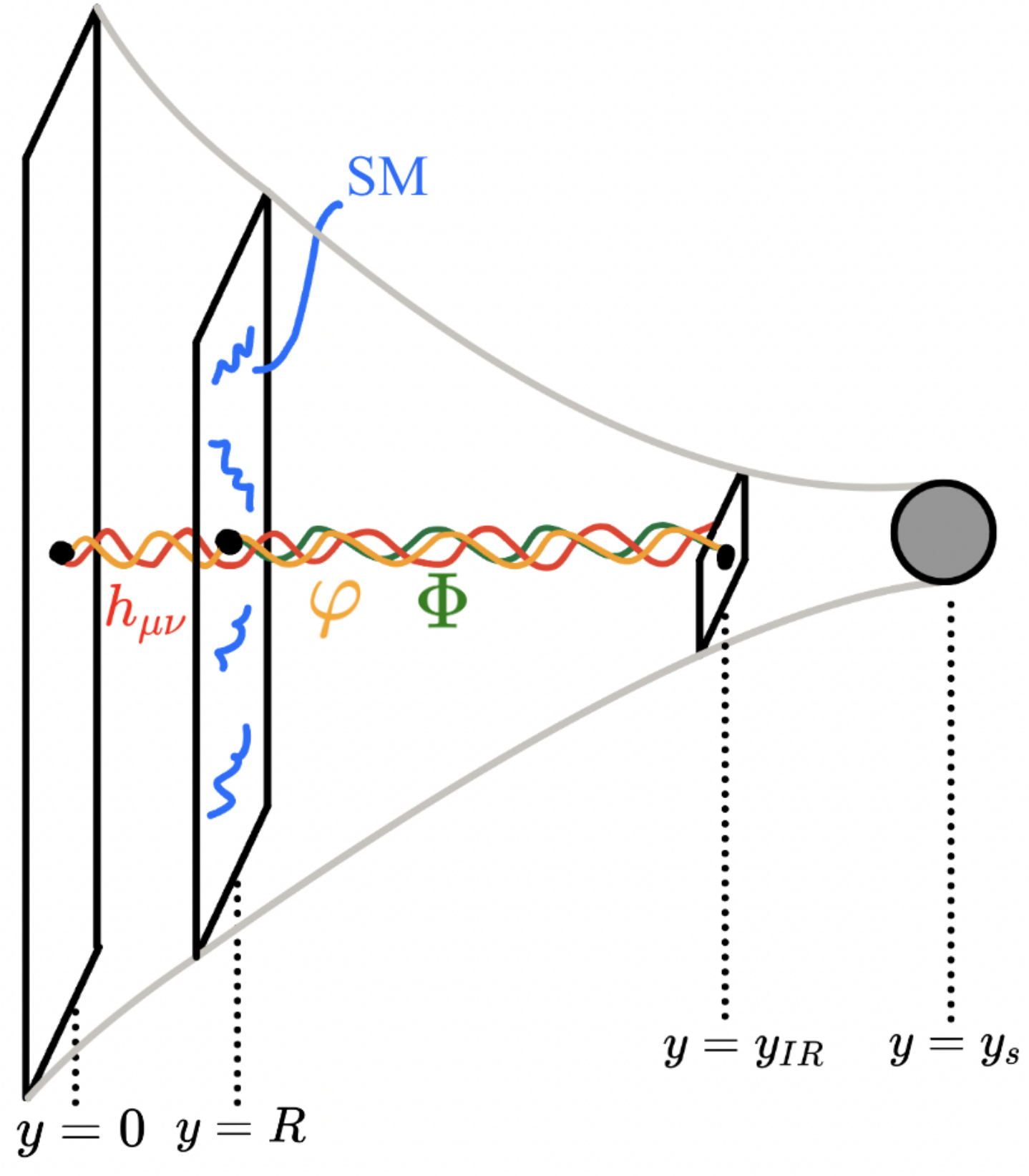}
\caption{5D setup of the near-continuum dark matter model.}
\label{fig:NearContinuumSetup}
\end{figure}

The semi-classical description of gravity encoded by the action~\leqn{5Daction} is only valid when all curvature invariants remain small compared to the 5D Planck scale. It will inevitably break down near the singularity, so the classical background in Eq.~\leqn{BG} ceases to be a good approximation as $y\to y_s$. A more complete theory of quantum gravity, such as string theory, is needed to provide a consistent description of physics in that region. It is expected that such a description will regulate the infinities that appear in Eq.~\leqn{BG}. While obtaining the soft-wall geometry from string theory is beyond the scope of this work, we can try to crudely model its effects by introducing an IR-regulator brane at $\yi $, close to $y_s$. This is schematically shown in Fig.~\ref{fig:NearContinuumSetup}. 

If the space ends on this brane, the singular region is excluded. To achieve this, the IR-brane action
is given by
\beq
S_{{\rm IR-reg}} \,=\, -\int d^4x \sqrt{-g_{\rm ind}} \gamma[\varphi(y=\yi)]\,,
\eeq{IRbrane}
where $\gamma=-\lambda=-6W$. The metric~\leqn{BG} satisfies the orbifold boundary conditions at $y=\yi$ with this brane action.  

In the presence of the IR regulator brane, KK decomposition of bulk fields will produce discrete towers, rather than gapped continuum spectra. However, the gapped continuum must reappear in the limit $\yi\to y_s$, so that if the regulator brane is placed sufficiently close to the singularity, the spacing between discrete KK levels $\Delta m$ is small compared to other relevant energy scales, in particular the gap scale:
\beq
\Delta m \ll \mu_0.
\eeq{near-cont}
We will refer to KK spectra satisfying this condition as {\it ``near-continuum"}. 

To find the KK mode spectrum, we first obtain the equations of motion for the scalar DM and graviton that follow from~\leqn{5Daction}. Writing the KK expansion of the scalar as 
\beq
\Phi(x, 0)=\sum_{n}\phi_{n}(x)f_{n}(y), 
\eeq{KKexp}
where the KK mode $\phi_{n}$ in the free theory satisfies $(\Box+m_{n}^{2})\phi_{n}=0$, the equation of motion for the profile $f_{n}(y)$ is
\begin{align}
    e^{-2A}(-f''_{n} + 4A'f'_{n})
    = 
    m_{n}^{2}f_{n}.
\end{align}
In conformally flat coordinates $z$, defined as $\frac{dz}{dy}=e^{A}$, this becomes 
\begin{align}
    -\ddot{f}_{n} + 3\dot{A}\dot{f}_{n}
    = 
    m_{n}^{2}f_{n},
\end{align}
where $\dot{(\,)}$ denotes a derivative with respect to $z$. We can turn this into a Schrödinger form by rescaling the profile $\psi_{n} = e^{-3A/2}f_{n}$. This yields
\begin{align}\label{eq:Schrodinger}
    -\ddot{\psi}_{n} + V(z)\psi_{n} 
    = 
    m_{n}^{2} \psi_{n},
\end{align}
where $V(z) = \frac{9}{4}\dot{A}^{2}-\frac{3}{2}\ddot{A}+m^2 e^{-2A}$. The graviton profile happens to satisfy the same differential equation (with $m=0$). The only differences between the graviton and scalar profile equations are their boundary conditions and the intervals in the 5th coordinate that each profile occupies. The graviton profile has support from $y=0$ to $y=y_{IR}$, while the scalar profile has support only from $y=y_{b}$ to $y=y_{IR}$. For the graviton profile (before the rescaling by $e^{-3A/2}$), we impose Neumann boundary conditions at $y=0$ and $y=y_{IR}$, while for the scalar profile $f(y)$ we impose Neumann at $y=y_{b}$ and Dirichlet at $y=y_{IR}$. These boundary conditions result in a zero-mode for the graviton (as required to reproduce 4D Newtonian gravity) and no zero-mode for $\Phi$ (as required to obtain cold/non-relativistic DM). 

\begin{figure}
    \centering
    \begin{subfigure}[c]{0.3\textwidth}
        \centering
        \includegraphics[width=\textwidth]{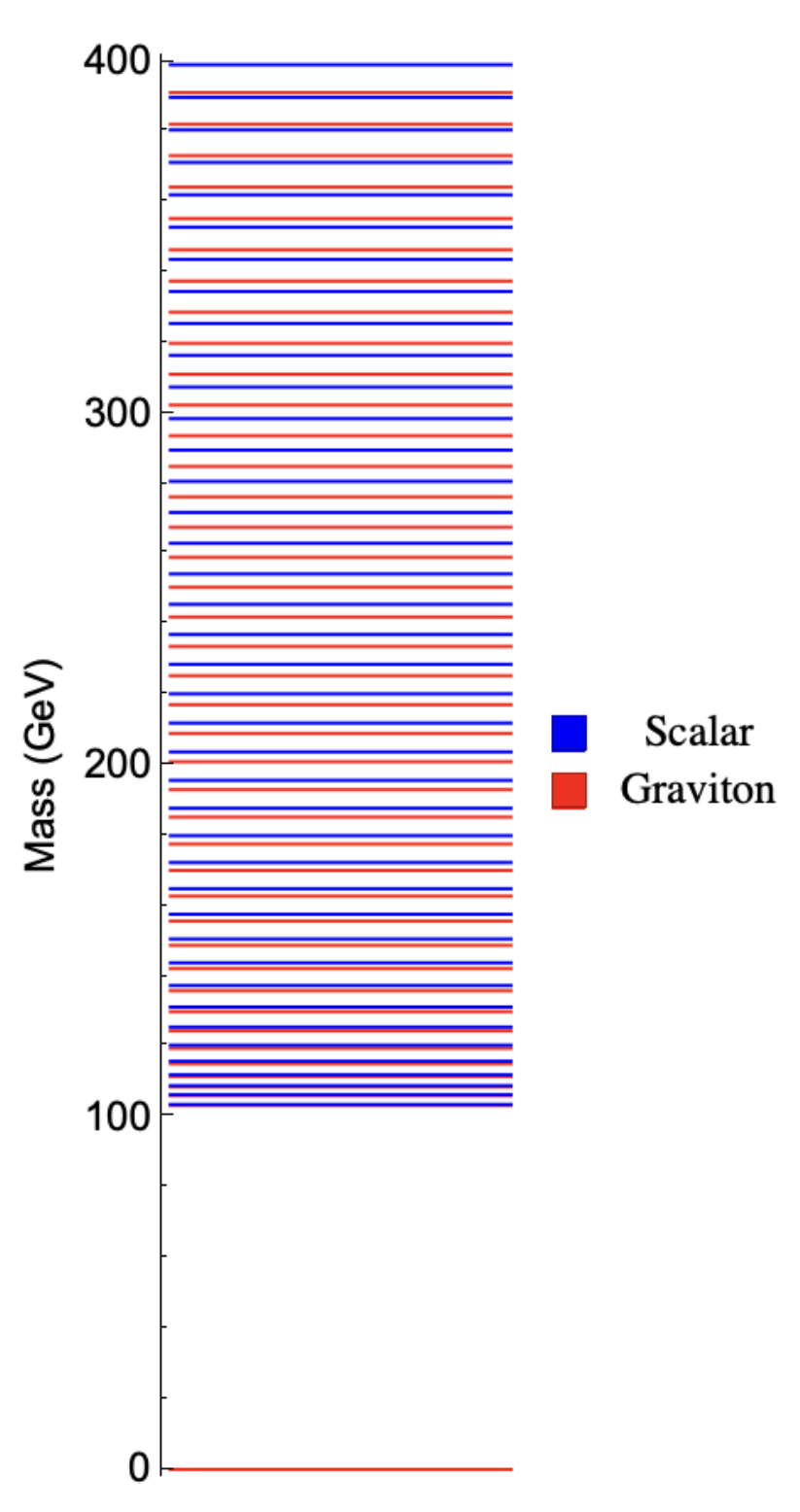}
        % \caption{}
    \end{subfigure}
    \hfill
    \begin{subfigure}[c]{0.6\textwidth}
        \centering
        \includegraphics[width=1.0\textwidth]{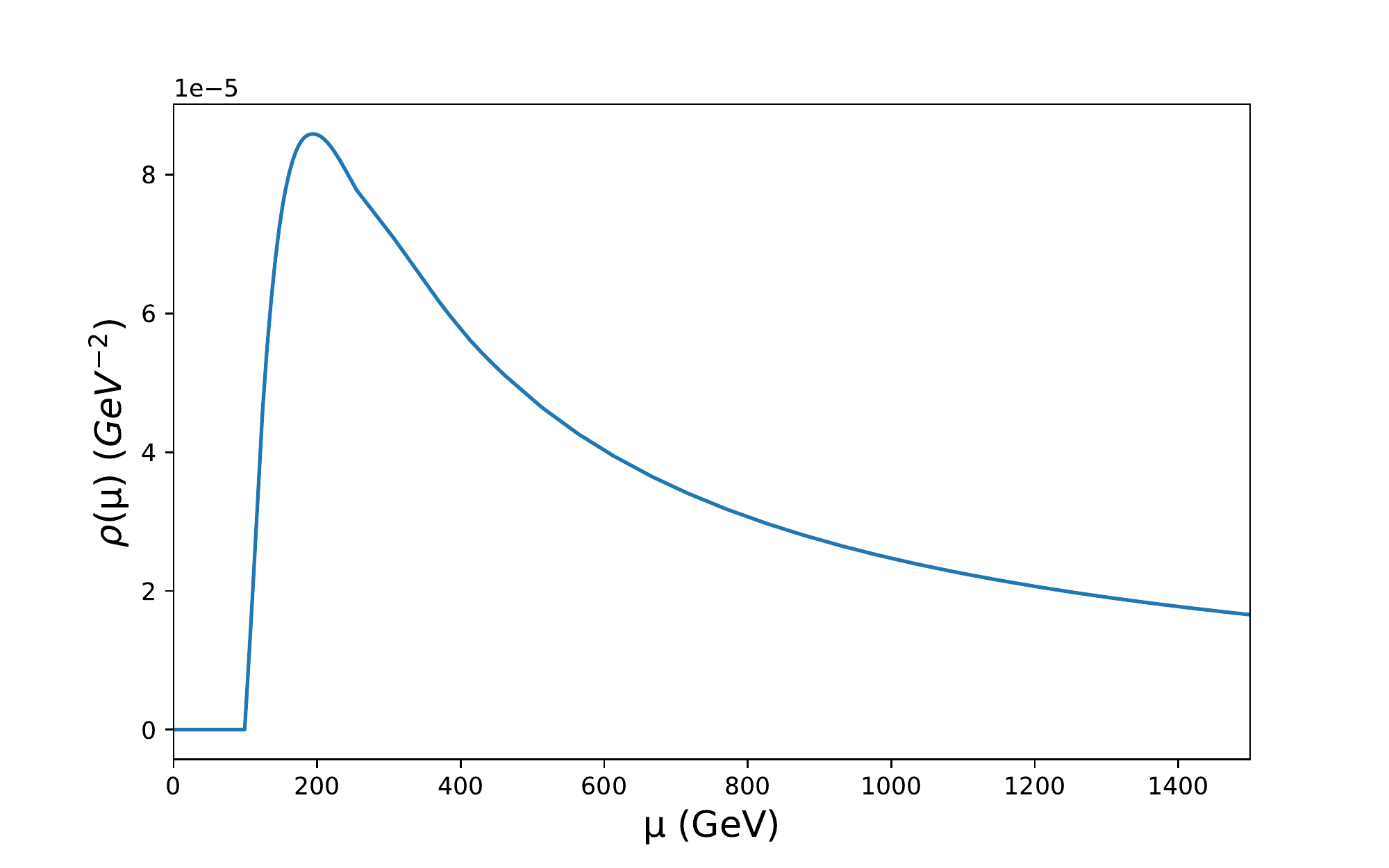}
        % \caption{}
    \end{subfigure}
\caption{Left panel: Spectra of the graviton and scalar DM KK towers in the benchmark near-continuum DM model. Right panel: Spectral density in the continuum limit of the same model.}
\label{fig:DiscreteSpectrum}
\end{figure}

To compute the masses for the graviton and the scalar KK modes, we numerically solve the eigenvalue problems defined by~\leqn{eq:Schrodinger} and the boundary conditions stated above. The resulting spectra are shown in Figure~\ref{fig:DiscreteSpectrum}. Since the graviton profile propagates on a larger volume, there is a sense in which it is `closer' to the continuum limit; the graviton mass spacings in comparison to the scalar DM mass spacings in Figure \ref{fig:DiscreteSpectrum} align with this intuition. 

The near-singularity region maps into $z\to \infty$ in the conformal coordinates, and $V(z)$ approaches a constant, $V\to \mu_0^2$, at large $z$. Thus the profiles $\psi(z)$ are approximately sinusoidal at large $z$, and the following scaling behaviors hold in the continuum limit  $z_{\rm IR} \to \infty$
\beq
\Delta m_n^2 = m_{n+1}^2-m_n^2 
\sim \frac{m_n}{z_{\rm IR}}\,,~~~~~~~
|\psi_n| \sim z_{\rm IR}^{-1/2}.
\eeq{estimates}
The second scaling is due to the normalization condition. In the calculation of physical observables, sums over KK states turn into integrals in the continuum limit. For example, the brane-to-brane propagator of the DM field is given by  
\beq
\langle \Phi(p, y_b) \Phi(-p, y_b) \rangle = \sum_n \frac{i|f_n(y_b)|^2}{p^2-m_n^2+i\varepsilon} \to \frac{1}{R} \int \frac{d\mu^2}{2\pi} \, \frac{i\,\rho(\mu^2)}{p^2-\mu^2+i\varepsilon}\,, 
\eeq{BtoB}
where $\rho$ is the spectral density, and $R=z(y_b)$. (The normalization of $\rho$ in this equation matches the choice made in Ref.~\cite{Csaki:2021gfm}.) This implies the following relationship between the parameters of the discrete theory and the spectral density of its continuum limit:
\beq
\lim_{\yi\to y_s} \frac{|f_n(y_b)|^2}{\Delta m_n^2} \,=\, \frac{1}{2\pi R} \, \rho(m_n^2). 
\eeq{limita}
Note that the finiteness of this limit is guaranteed by the scalings in Eq.~\leqn{estimates}. Using this relationship, sums over KK states in cross sections and decay rates in a near-continuum theory can be approximated by continuous mass integrals, weighted by the spectral density:
\beq
R \, \sum_n |f_n(y_b)|^2 \,\to\,\int \frac{d\mu^2}{2\pi}\,\rho(\mu^2).
\eeq{XS_limit}
This matches the prescription for the calculation of inclusive cross sections with continuum in the final state used in Refs.~\cite{Csaki:2021gfm,Csaki:2021xpy}.

In a free theory, each KK mode behaves as an independent 4D field. For example, the two-point function for the KK modes $\phi_k(x)$ of the DM field is 
\beq
\langle \phi_k(p) \phi_l(-p)\rangle \,=\, \frac{i\delta_{kl}}{p^2-m_k^2+i\varepsilon}.
\eeq{prop}
When interactions of the DM field (with the SM, on the brane, as well as with gravitons, in the bulk) are included, loop corrections induce a self-energy correction to the propagator~\leqn{prop}, $\Pi_{kl}(p^2)$. The real part of $\Pi$ shifts the physical masses of the KK modes, while the imaginary part describes their decays. Depending on the model parameters, two situations are possible:

\begin{enumerate}

\item Im\,$\Pi_{kl} \ll |m_k^2-m^2_{k\pm 1}|$, for any $k$ and $l$. In this case, narrow-width approximation (NWA) is applicable for each KK mode. Each mode can be treated as an ordinary 4D particle when calculating the S-matrix. Sums over KK modes in cross section and decay width calculations are performed at the level of matrix element-squared, {\it i.e.} there is no interference among them.   

\item Im\,$\Pi_{kl} \gsim |m_k^2-m^2_{k\pm 1}|$. Here NWA breaks down, and the original KK modes no longer act as asymptotic states in the calculation of the S-matrix.  

\end{enumerate}

The main benchmark model studied in this paper is in the first category. This allows for calculation of detailed properties of near-continuum DM production and decay at colliders, including particle multiplicity, energy spectra, angular distributions, etc. The benchmark is defined by the following parameters:
\beqa
k&=&10^{10}~{\rm GeV};~~\mu_0=100~{\rm GeV};~~R^{-1}=80~{\rm GeV};~~M_5=2\cdot 10^{16}~{\rm GeV};\CR y_{\rm IR} &=&1.6\cdot 10^{-9}~ {\rm GeV^{-1}}; ~~~\sin^2\alpha\,=\,0.1
\eeqa{BM_model}
Note that the parameters are chosen such that the known 4D Planck scale is reproduced, while the gap scale is around the weak scale as required in the $Z$-portal continuum DM model. The regulator brane location determines the splitting $\Delta m_n$ between the neighboring KK modes; for the model defined by Eq.~\leqn{BM_model}, there are 40 KK states with masses between 100 and 400 GeV. The KK spectrum of this model, along with the spectral density obtained in the $\yi\to\infty$ limit, are illustrated in Fig.~\ref{fig:DiscreteSpectrum}. The calculation of decay widths of each KK mode is discussed in the next section, where it will be demonstrated that the NWA condition is satisfied, and the DM decays occur mainly via emission of (on- or off-shell) $Z$ bosons.

\section{Production and Decays of Near-Continuum DM}
\label{sec:NCProdDecay}

In this section, we discuss production and decay of near-continuum DM states at colliders. As a concrete example, we consider electron-positron collisions at $\sqrt{s}=500$ GeV, as may be provided in the future by the proposed International Linear Collider (ILC)~\cite{ILC:2013jhg,ILCInternationalDevelopmentTeam:2022izu} or Cool Copper Collider (C3)~\cite{Dasu:2022nux}. We expect that qualitative phenomenological features uncovered by our analysis will be also applicable at hadron colliders, such as the LHC. A detailed analysis of the LHC phenomenology will be the subject of future work. 

\subsection{Near-Continuum DM Production}
\label{subsec:NCProduction}

\begin{figure}
\centering
\includegraphics[width=5.9in]
{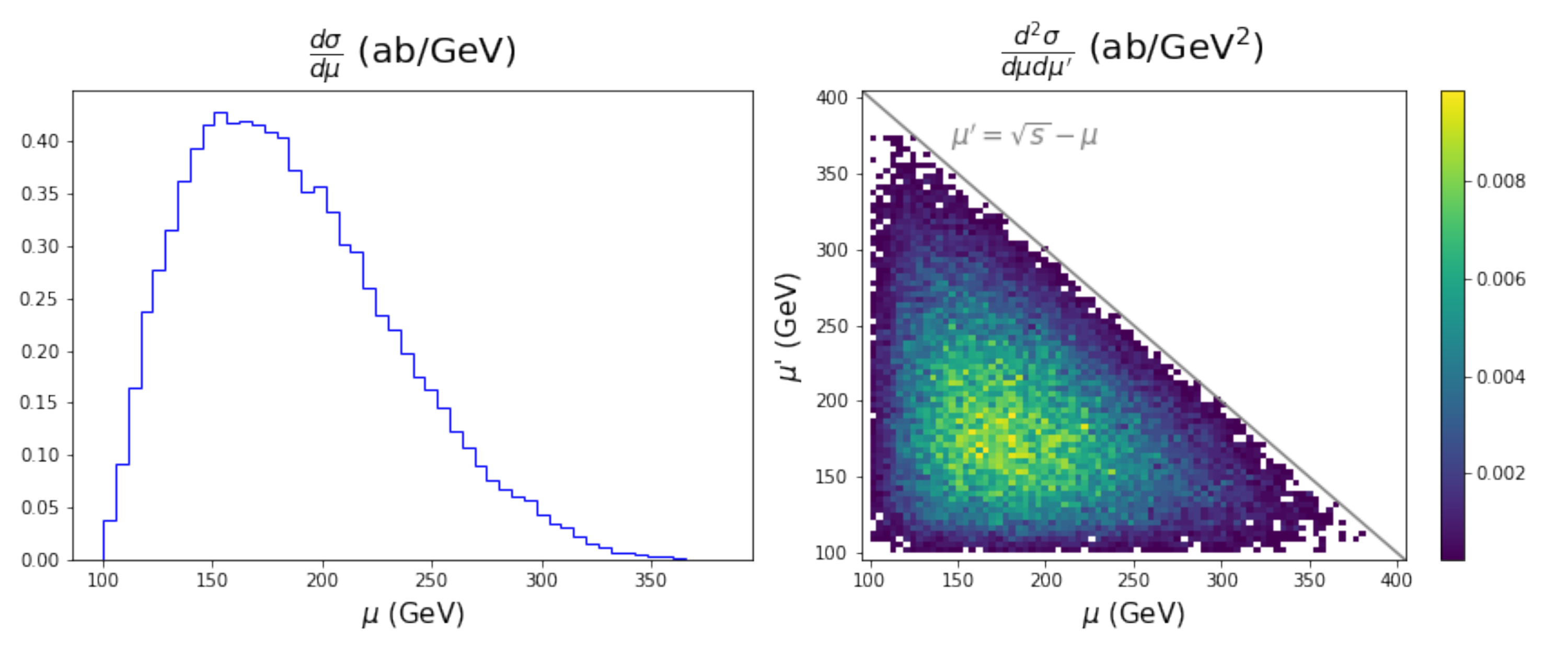}
\caption{1D (left) and 2D (right) distributions of masses of the near-continuum DM states produced in $e^+e^-$ collisions at $\sqrt{s}=500$ GeV.}
\label{fig:VegasProductionMasses}
\end{figure}

\begin{figure}
\centering
\includegraphics[width=3.5in]
{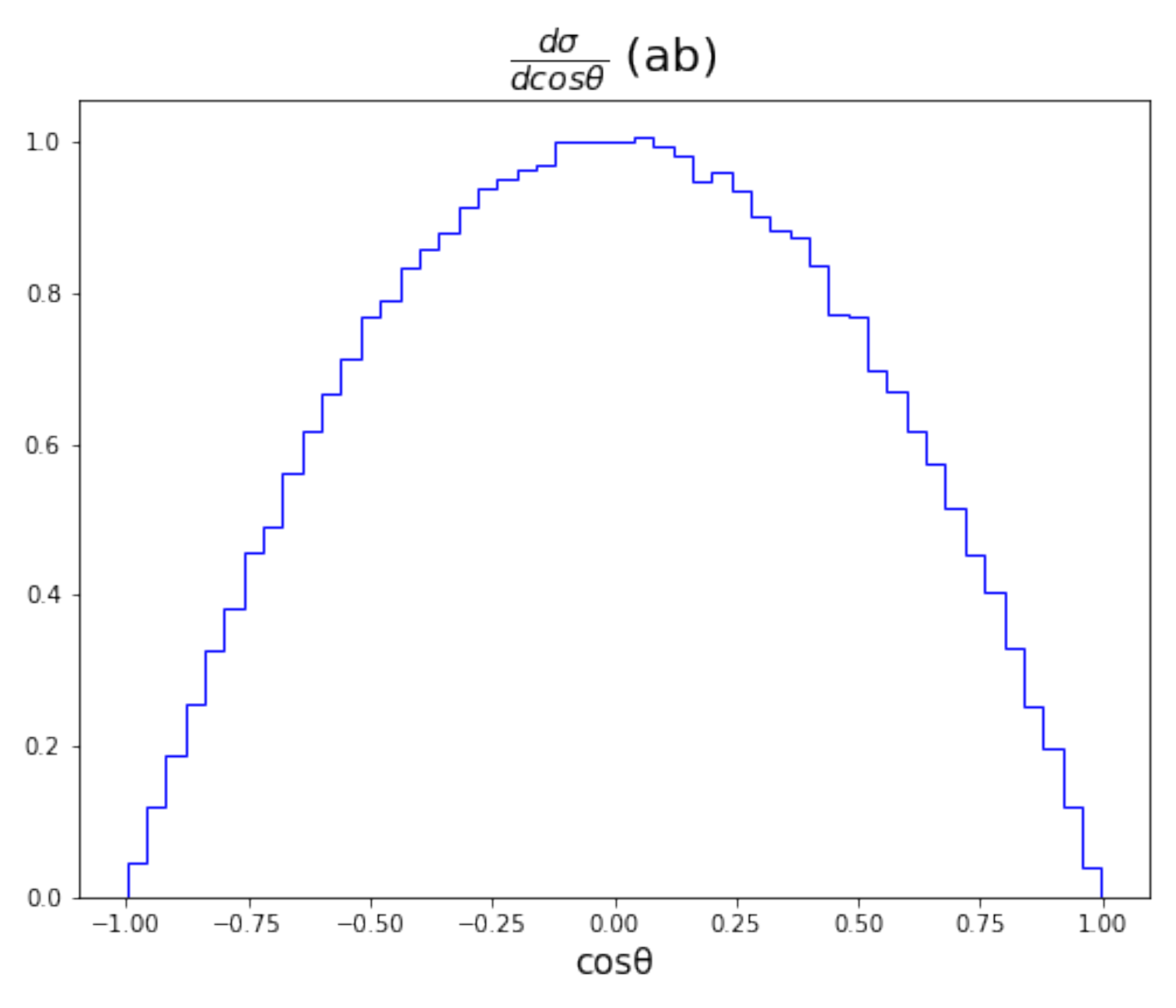}
\caption{Angular distribution of the near-continuum DM states produced in $e^+e^-$ collisions at $\sqrt{s}=500$ GeV.}
\label{fig:VegasProductionAngles}
\end{figure}

Near-continuum DM states are produced in the $s$-channel via their coupling to $Z$ bosons. All KK states are odd under the DM $Z_2$ symmetry, while the SM particles are even, so that the KK states must be pair-produced in SM collisions. At tree level, the only diagram is
\begin{align}
    \feynmandiagram [inline=(a.base), horizontal=a to b, large] {
i1 [particle=\(e^{-}\)]  -- [fermion] a -- [fermion] i2  [particle=\(e^{+}\)]  ,
a -- [boson, edge label = \(Z\)] b,
f1  [particle=\(\phi(\mu)\)] -- b -- f2 [particle=\(\phi(\mu')\)],
}; .
\end{align}
In principle, this process can be simulated using conventional Monte Carlo (MC) generators, such as {\tt MadGraph}~\cite{Alwall:2011uj}. In practice, however, this is not feasible. Since momentum in the fifth dimension is not conserved, any pair of the KK modes, within kinematic constrains, may be produced. In our benchmark model and with $\sqrt{s}=500$~GeV, this results in $40\times 40=1600$ possible final states. Each of these states in turn will undergo cascade decays, further enhancing the number of ultimate final states. Instead, we take an alternative route to numerical simulation of this process. Using Eq.~\leqn{XS_limit}, we can approximate the inclusive DM production cross section as  
\beq
    \sigma = 
    \frac{\text{sin}^{4}\alpha \, g_{Z}^{2} \, c_{e}}
         {32\pi\sqrt{s}(s-m_{Z}^{2})^{2}}
         \int_{\mu_{0}}^{\sqrt{s}-\mu_{0}}
         \frac{d\mu^{2}}{2\pi} \, 
         \rho(\mu)
         \int_{\mu_{0}}^{\sqrt{s}-\mu}
        \frac{d\mu'^{2}}{2\pi}
         \, 
         \rho(\mu') 
         \,\int_{-1}^{1}
         d\text{cos}\theta \,\,
         \frac{d\sigma_{\mu\mu'}}{d\cos\theta}\,,
\eeq{eq:ProdXSec}
where 
\beq
c_{e} = \frac{g^{2}}{2\text{cos}^{2}\theta_{w}}\sum_{i}(T^{3}_{i}-\text{sin}^{2}\theta_{w} Q_{i})^2~~~~~~~~~~~~~(i=e_{L}, e_{R}).
\eeq{ce_def}
The differential cross section can be written in terms of sums and differences of the squared masses $\Sigma\equiv \mu^{2}+\mu'^{2}$ and $\Delta\equiv \mu^{2}-\mu'^{2}$: 
\beq
   \frac{d\sigma_{\mu\mu'}}{d\cos\theta} = 
   \bigg(
   \frac{s^{2}-2s\Sigma+\Delta^{2}}{4s}
   \bigg)^{3/2}\,(1-\cos^2\theta). 
\eeq{diff_xsec}
We then use {\tt Vegas}~\cite{Lepage:1977sw} to perform the integrals in Eq.~\leqn{eq:ProdXSec}, and to sample the distribution of the produced DM particle masses and scattering angles. (For details, see Appendix.) The total cross section for our benchmark model is
\beq
\sigma(e^+e^-\to \phi\phi) = 67~{\rm ab}\,,
\eeq{sigma_prod}
yielding a substantial sample of DM events for a typical projected ILC integrated luminosity of a few ab$^{-1}$. Mass distributions are shown in Fig.~\ref{fig:VegasProductionMasses} and the angular distribution is shown in Fig.~\ref{fig:VegasProductionAngles}. As expected, the shape of the mass distributions roughly mirrors the spectral density, while the angular distribution is characteristic of scalar production in fermion collisions mediated by a vector boson.    

\subsection{Near-Continuum DM Decay}
\label{subsec:NCDecay}

After production, a DM state of mass $\mu_k>\mu_{0}$ will decay to another DM state of mass  $\mu_l\in (\mu_{0}, \mu_{k})$. Two decay channels are available in our model:
\beq
\phi_k \to \phi_l + Z^{(*)} \to \phi_l + f\bar{f};
\eeq{decZ}
\beq
\phi_k \to \phi_l + G_m.
\eeq{decG}
Here $Z^{(*)}$ is an SM $Z$ boson which may be on- or off-shell depending on the DM masses, $f$ denotes any of the kinematically accessible SM fermions, and $G_m$ denotes any kinematically accessible KK graviton (including the zero mode, $m=0$). Note that the presence of a DM KK mode $\phi_l$ in the final state is required by the conserved $Z_2$ symmetry. Below we will evaluate the partial width of each of the decay modes. We find that for the benchmark model defined in Eq.~\leqn{BM_model}, decays into SM fermion pairs dominate. Following this decay, the produced DM state $\phi_l$ will itself decay into another, lighter DM state and a pair of SM fermions, and so on, resulting in a ``cascade decay" event topology. These cascades and resulting phenomenological signatures will be considered in Section~\ref{sec:VisiblePheno}.    

\subsubsection{Z-portal Decay}
\label{subsubsec:ZDecay}

The decay~\leqn{decZ} is described by the diagram 
\begin{align}
\begin{tikzpicture}
\begin{feynman}
\vertex (a) {\(\phi(\mu_{k})\)};
\vertex [right=of a] (b);
\vertex [above right=of b] (f1) {\(\phi(\mu_{l})\)};
\vertex [below right=of b] (c);
\vertex [above right=of c] (f2) {\(\bar{f}\)};
\vertex [below right=of c] (f3) {\(f\)};
\diagram* {
(a) -- (b) -- (f1),
(b) -- [boson, edge label'=\(Z\)] (c),
(c) -- [anti fermion] (f2),
(c) -- [fermion] (f3),
};
\end{feynman}
\end{tikzpicture}.
\end{align}

\begin{figure}
\centering
\includegraphics[width=5.8in]{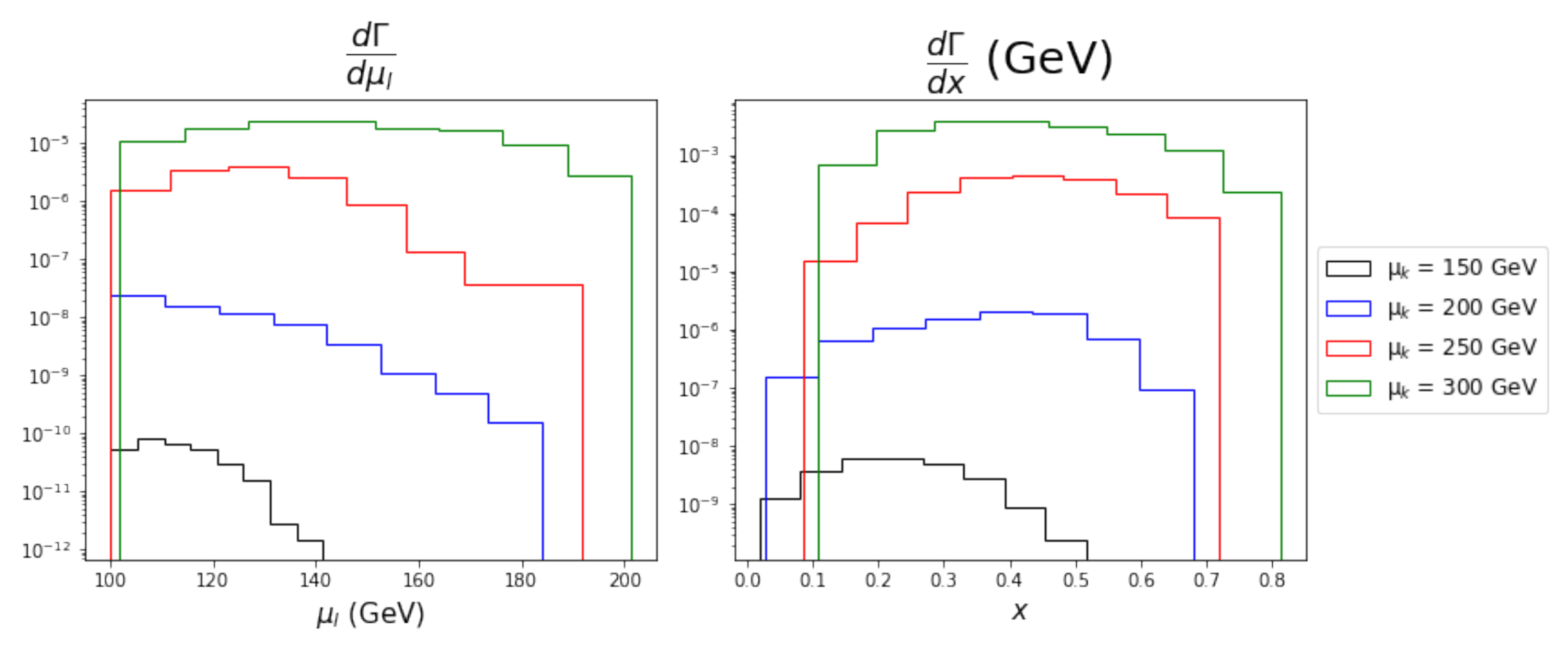}
\caption{\label{fig:epsart} Distribution of the final-state DM mass (left) and the fermion energy fraction (right) in the decay~\leqn{decZ}, with fixed $\mu_{k}$ values of 150, 200, 250, and 300 GeV.}
\label{fig:VegasDecay}
\end{figure}

\noindent In the rest frame of $\phi(\mu_{k})$, the rate for this decay is given by
\begin{align}
    \Gamma_{Z} &= 
    \frac{\text{sin}^{4}\alpha \, g_{Z}^{2}c_{f}}{8\mu_{k}} \,
    [ \rho(\mu_{k})\Delta \mu^{2} ]
    \int
    % _{\mu_{0}}^{\mu_{1}} 
    \frac{d\mu_{l}^{2}}{2\pi}
    \rho(\mu_{l})
    \int d\Pi_{3}
    \,\, 
    \Gamma_{\mu_{l}x_{f}x_{\bar{f}}}\,.
\end{align}
Here the term in the square brackets is the value of the DM field profile on the SM brane, as approximated by Eq.~\leqn{limita}, and $\Delta\mu^2=\mu_{k+1}^2-\mu_k^2$. The coefficient $c_{f}$ is the square of the coupling of the $Z$ boson to SM fermions:
\begin{align}
    c_{f} = 
    \frac{g^{2}}{2\text{cos}^{2}\theta_{w}} 
    \sum_{i}(T^{3}_{i} - \text{sin}^{2}\theta_{w} Q_{i})^{2}\,,
\end{align}
where $i$ runs over all kinematically accessible SM fermions.  
The 3-body phase space integral is given by (defining $R\equiv \mu_{l}^{2}/\mu_{k}^{2}$)
\begin{align}
    \int d\Pi_{3} = 
    \frac{\mu_{k}^{2}}{128\pi^{3}} 
    \int_{0}^{1-R} dx_{f} 
    \int_{1-R-x_{f}}^{\frac{1-R-x_{f}}{1-x_{f}}} dx_{\bar{f}},
\end{align} 
and the differential decay rate takes the form
\begin{align}
    \Gamma_{\mu_{l}x_{f}x_{\bar{f}}} = 
    8\mu_{k}^{4}
    \frac{1+R+x_{f}x_{\bar{f}}-x_{f}-x_{\bar{f}}} 
         {|\mu_{k}^{2}(x_{f}+x_{\bar{f}}+R-1) - m_{Z}^{2} + i\Gamma_{Z}m_{Z}|^{2}}\,,
\end{align}
where $x_{f,\bar{f}}\equiv2E_{f,\bar{f}}/\mu_{k}$ are the energy fractions of the fermions, which are assumed to be massless. Note that the equations above allow for a unified treatment of on- and off-shell $Z$ bosons, which turns out to be numerically feasible in this case. 
Once again, we use {\tt Vegas} to perform the integrals and sample from the distributions in the decay product phase space. Distributions of the final-state DM mass and the fermion energies, for a fixed decaying DM mass, are shown in Fig.~\ref{fig:VegasDecay}.   

It is interesting to note that the decay rate of a KK mode through the $Z$-portal is proportional to $\Delta\mu^2$, or equivalently to the value of the KK wavefunction on the SM brane $f(y_b)$, both of which go to zero in the continuum limit. This implies that if this were the only possible decay channel, the narrow-width approximation (NWA) for each KK mode could continue to be satisfied even in the continuum limit. Decays involving KK gravitons, where the relevant couplings receive contributions throughout the bulk, generally violate NWA in the continuum limit. For near-continuum models (with finite $z_{\rm IR}$), validity of the NWA depends on the other model parameters, as will be discussed in the next section.     

\subsubsection{KK Graviton Decay}
\label{subsubsec:GravDecay}

The decay~\leqn{decG} is described by the diagram 
\begin{align}
% \resizebox{150}{120}{
\begin{tikzpicture}
\begin{feynman}
\vertex (a) {\(\phi(\mu_{k})\)};
\vertex [right=of a] (b);
\vertex [above right=of b] (f1) {\(\phi(\mu_{l})\)};
\vertex [below right=of b] (c)  {\(h^{m}_{\mu\nu}\)};
\diagram* {
(a) -- (b) -- (f1),
(b) -- [graviton] (c),
};
\end{feynman}
\end{tikzpicture}
% }
\end{align}
The corresponding decay rate is
\beq
    \Gamma_{g}\,=\, \sum_{l,m}
    \frac{1}{2\mu_{k}}
    \bigg|
    \int_{R}^{z_{\rm IR}}
    e^{-3A(z)}
    h_{m}(z)f_{k}(z)f_{l}(z)
    dz
    \bigg|^{2}
    \int d\Pi_{2}
    \sum_{\text{spins}}
    |\mathcal{M}_{g}|^{2}\,,
\eeq{decGrate}
where $h_m(z)$ is the normalized profile of the graviton KK mode.
The spin-averaged matrix element squared is
\beqa
    \sum_{\text{spins}}
    |\mathcal{M}_{g}|^{2}
    &=& 
    \frac{1}{M_{5}^{3}}
    p_{k}^{\mu}p_{l}^{\nu}
    p_{k}^{\alpha}p_{l}^{\beta}
    P_{\mu\nu\alpha\beta}(p_{m})
    \CR 
    &=& 
    \frac{
    (\mu_{l}^2-m_{m}^{2})^{2}
    (
    (\mu_{l}^{2}-\mu_{k}^{2})^{2}
    + m_{m}^{4}
    - 2m_{m}^{2}(\mu_{l}^{2}+\mu_{k}^{2})
    )^{2}
    }
    {24 M_{5}^{3} m_{m}^{4}\mu_{k}^{4}},
\eeqa{MG}
where $p_{k,l,m}$ are the 4-momenta of the KK modes, $m_m$ is the KK graviton mass, and $M_{5}$ is the 5D Planck scale. The function $P_{\mu\nu\alpha\beta}$ is the numerator of the massive graviton propagator. 

For $k\sim M_5$, the graviton decay rate calculated according to Eqs.~\leqn{decGrate},~\leqn{MG} is generically larger than the mass splitting between the neighboring KK modes, indicating breakdown of the narrow-width approximation. In this regime, perturbative calculations using the KK states as asymptotic states are not applicable. However, as $k$ is decreased, $\Gamma_g$ decreases, and model parameters where perturbation theory applies can be found. To understand this, consider varying $k$, and adjusting the other model parameters so that the 4-dimensional gravitational scale $M_4$, the gap scale $\mu_0$, and the splitting between the neighboring KK modes $\Delta m_k^2$ are all fixed. Several factors determine the scaling of the graviton decay rate $\Gamma_{g}$. First, it is  proportional to $M_{5}^{-3}$, which roughly scales like $1/k$ according to Eq.~\leqn{MPl4D}. Second, the graviton decay rate is also proportional to the overlap integral
\begin{align}
\bigg|
    \int_{R}^{z_{IR}} e^{-3A} 
    f_{i}(z)
    f_{m}(z)
    h_{n}(z)
    dz 
\bigg|^{2} \,.
\end{align}
The bounds of the integral remain essentially unchanged in the $z$ coordinates. The warp factor scales roughly as $e^{-3\text{log}(kz)} \sim k^{-3}$. A normalized KK profile scales as $f(z) = \hat{f}(z)/\sqrt{\int e^{-3A}\hat{f}^{2}dz} \sim k^{3/2}$. Putting it all together, $\Gamma_{g}$ scales roughly as $k^{-1}(k^{-3}k^{9/2})^{2} \sim k^{2}$. At the same time, the DM decay rate through the $Z$ portal is approximately independent of $k$ in this scaling. Thus, for sufficiently low value of $k$, we expect that the graviton decay rate will not only become much smaller than the KK mass splitting, but also will be subdominant to the $Z$-portal decay rate. This is the case for the benchmark parameters, Eq.~\leqn{BM_model}, used in our phenomenological study.

\section{Phenomenology: DM Cascade Decays}
\label{sec:VisiblePheno}

As described above, the near-continuum DM states are pair-produced in $e^+e^-$ collisions, and each DM state then undergoes a cascade decay. A lighter DM state, along with a pair of SM fermions, are produced at each step in the cascade decay. Since the lifetime of a DM state increases with decreasing mass, the cascade effectively terminates when a DM state with sufficiently long lifetime to escape the detector is reached. The resulting phenomenological signature is missing energy (ME) from the escaping DM states, plus multiple SM fermions from cascade decays. To study this signature in more detail, we have constructed a custom-made Monte Carlo (MC) simulation tool, described in more detail in Appendix~\ref{app:MCFramework}. The tool uses {\tt VEGAS}~\cite{Lepage:1977sw} to simulate DM pair-production and the DM decay at each step in the cascade in the DM rest frame, and then boosts the four-momenta of the decay products to obtain lab-frame distributions of the final-state particles. Using this tool, we have simulated the production and decay of near-continuum DM, within the benchmark model introduced in section~\ref{sec:model}, in $e^+e^-$ collisions at $\sqrt{s}=500$~GeV. In this section, we present the results of this simulation.

\begin{figure}
\centering
\includegraphics[width=2.8in]{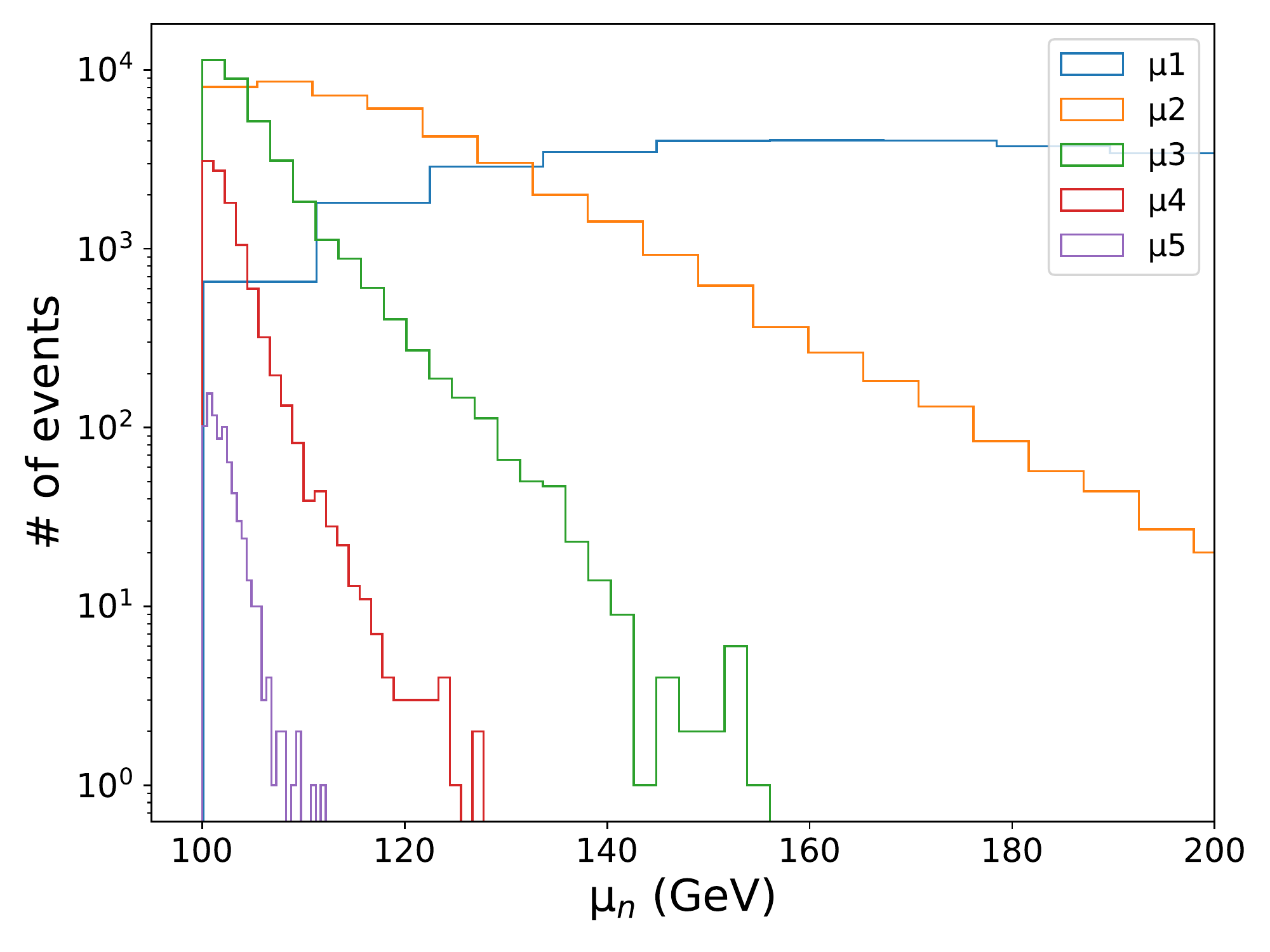}
\includegraphics[width=2.8in]{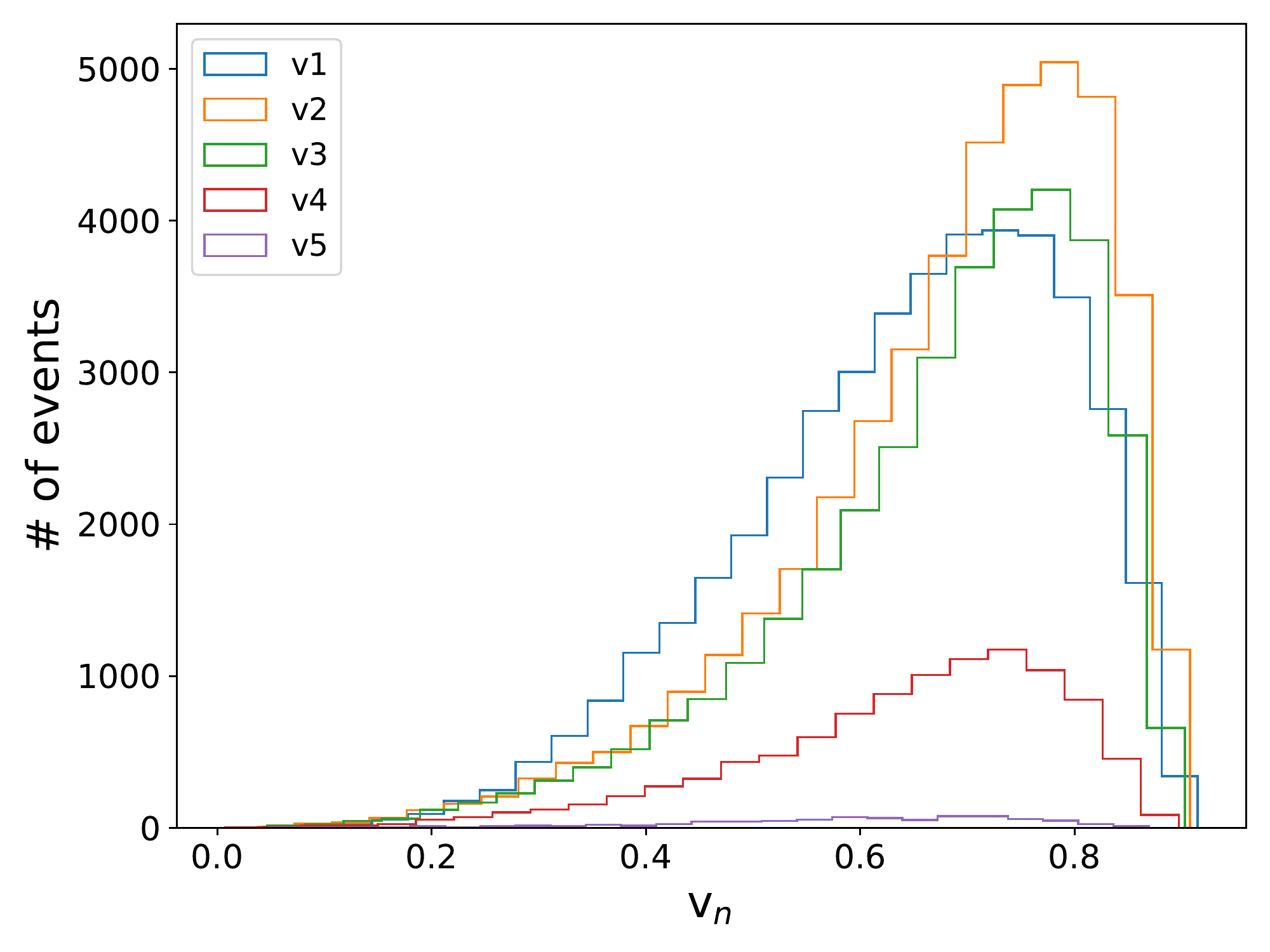}
\caption{Distributions of the DM state mass (left) and lab-frame velocity (right), for the first 5 steps in the cascade.}
\label{fig:masses}
\end{figure}

Figure~\ref{fig:masses} illustrates mass and lab-frame velocity distributions of the DM states produced in the cascade. The distributions labeled $\mu_1$ and $v_1$ are those of the originally pair-produced DM, while those labeled $\mu_i$ and $v_i$ with $i=2\ldots 5$ correspond to DM particles produced in the first four steps of the ensuing cascade. While not directly observable, these distributions are useful for understanding the kinematics of the signal events. As expected, the DM masses steadily decrease, while lab velocities increase, with each subsequent decay. In our simulation, we model the effect of the finite detector volume by terminating each cascade when the lab-frame decay length of the DM particle exceeds 327~cm (corresponding to $\mu\leq 106$~GeV for a DM with velocity 0.65$c$, typical of our signal), since such particles would decay outside of the detector. This accounts for decreasing number of events with each cascade step, clearly visible in Fig.~\ref{fig:masses}. While not shown in this figure, our sample includes events with $>4$ steps in the cascade, as long as the decay-length condition is satisfied.    

\begin{figure}
\centering
\includegraphics[width=2.8in]{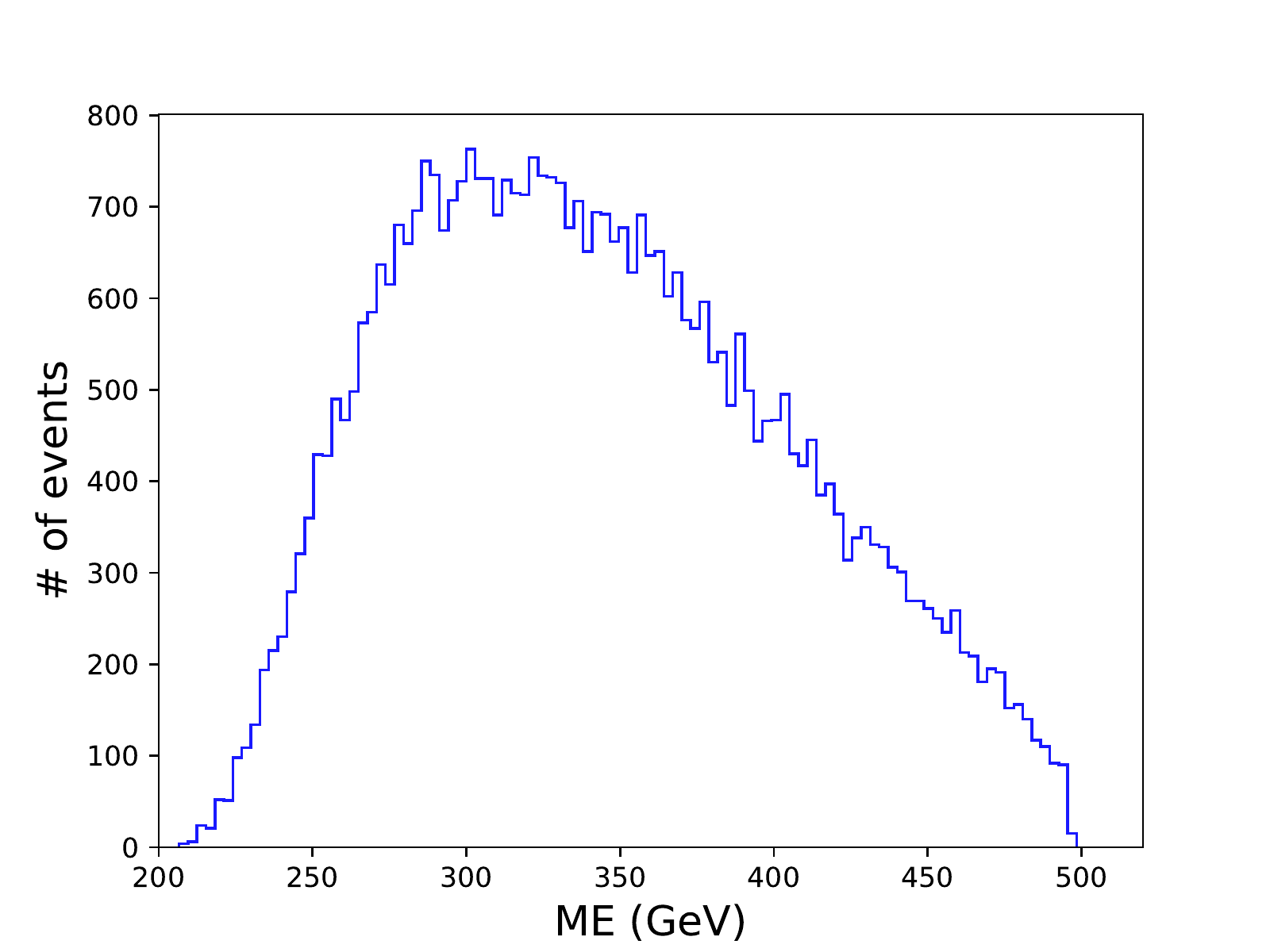}
\includegraphics[width=2.8in]{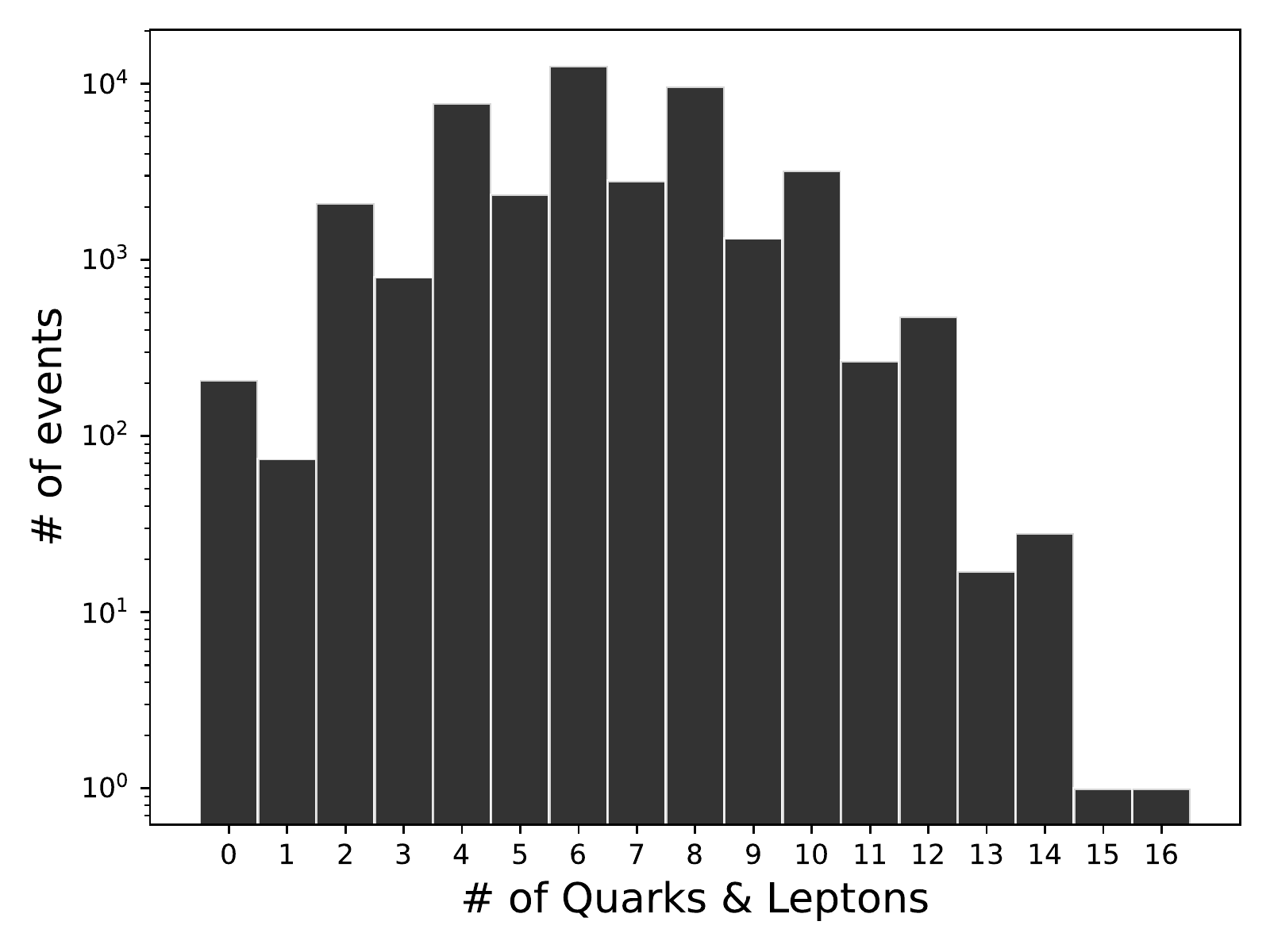}
\caption{Missing energy (left) and fermion multiplicity (right) distributions in events with near-continuum DM production. Fermions passing a minimum energy cut $E\geq 1$~GeV are included.}
\label{fig:MET}
\end{figure}
\begin{figure}

\centering
\includegraphics[width=2.8in]{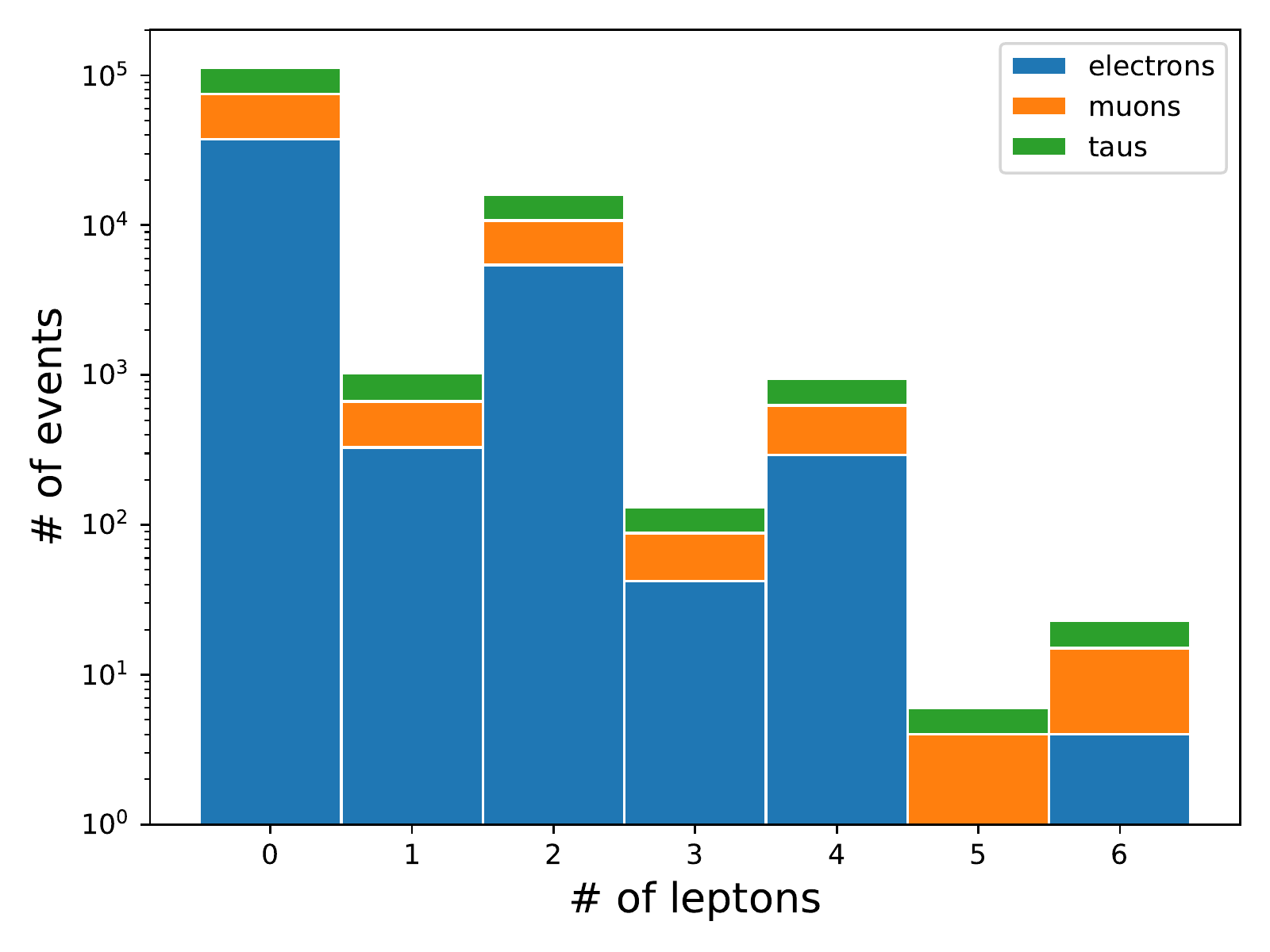}
\includegraphics[width=2.8in]{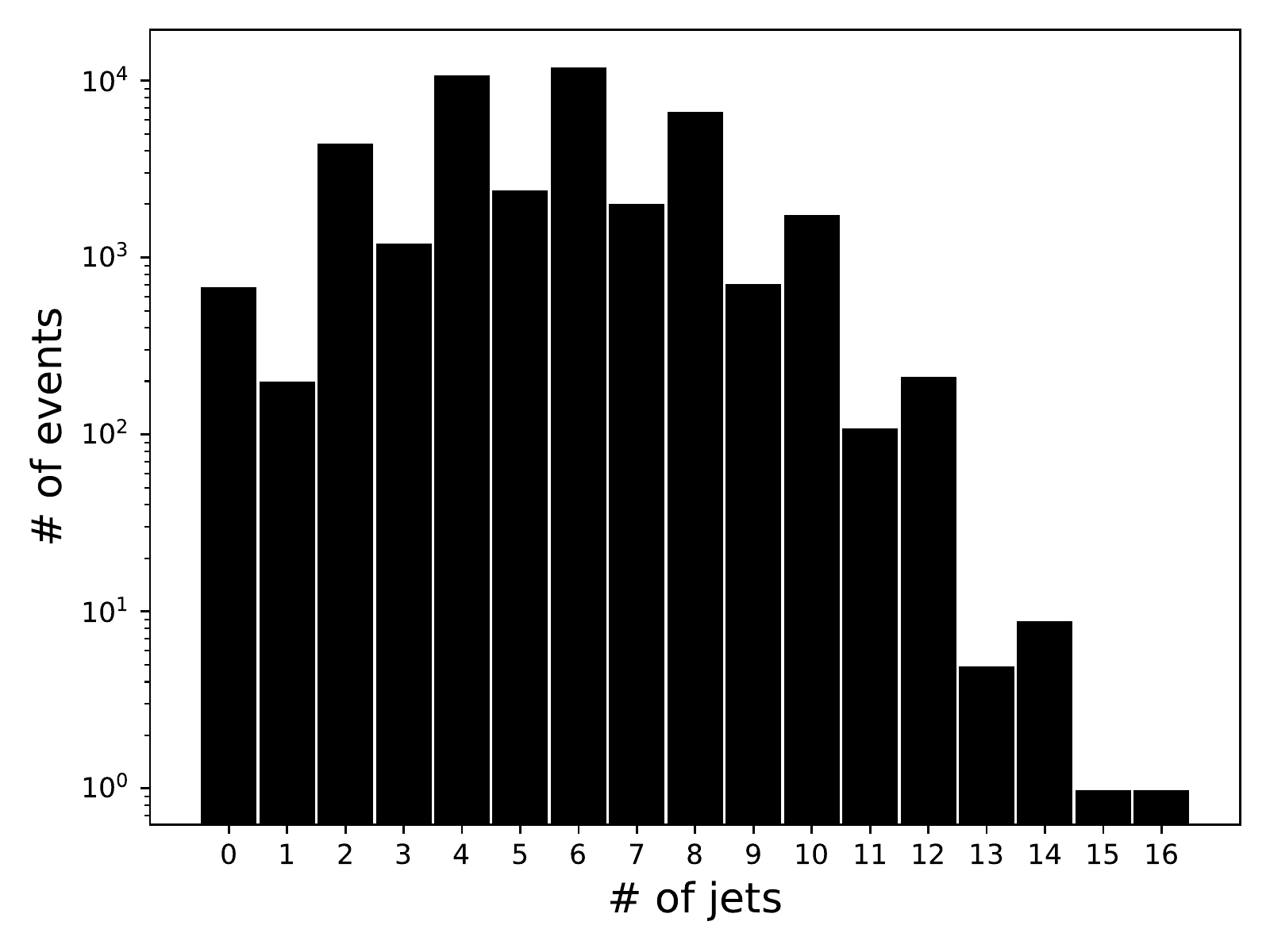}
\caption{Lepton multiplicity (left) and jet multiplicity (right) distributions in events with near-continuum DM production. Leptons and jets passing a minimum energy cut $E\geq 1$~GeV are included.}
\label{fig:Mult}
\end{figure}

The characteristic signature of near-continuum DM production events are significant missing energy and high multiplicity of SM fermions from cascade decays. These two observables are shown in Fig.~\ref{fig:MET}. In the left panel, only events with ME < 499 GeV are shown, and in the right panel, only charged leptons and quarks with energies above 1 GeV are included.\footnote{Since two fermions are produced in each DM decay, the fermion multiplicity is typically even; odd multiplicities are only generated when one of the fermions fails the minimum energy cut. This accounts for the ``sawtooth" structure in the multiplicity plots in Figs.~\ref{fig:MET} and~~\ref{fig:Mult}.} The energy carried by neutrinos is included in the left panel, although their contribution to missing energy is subdominant to that of the DM. A typical event has 300-350 GeV of missing energy, and as many as 8 SM fermions. Both numbers are well above expectations for SM processes in $e^+e^-$ collisions, and it should be straightforward to separate the DM signal from the SM backgrounds using cuts on these observables. 
In each decay, the branching ratios of lepton and quark final states are determined by the SM $Z$ couplings; in particular, if the available energy is well above the $b$ quark mass, these branching ratios are identical to those of the on-shell $Z$. The resulting multiplicities of leptons and jets are shown in 
Fig.~\ref{fig:Mult}.

\begin{figure}
\centering
\includegraphics[width=5in]{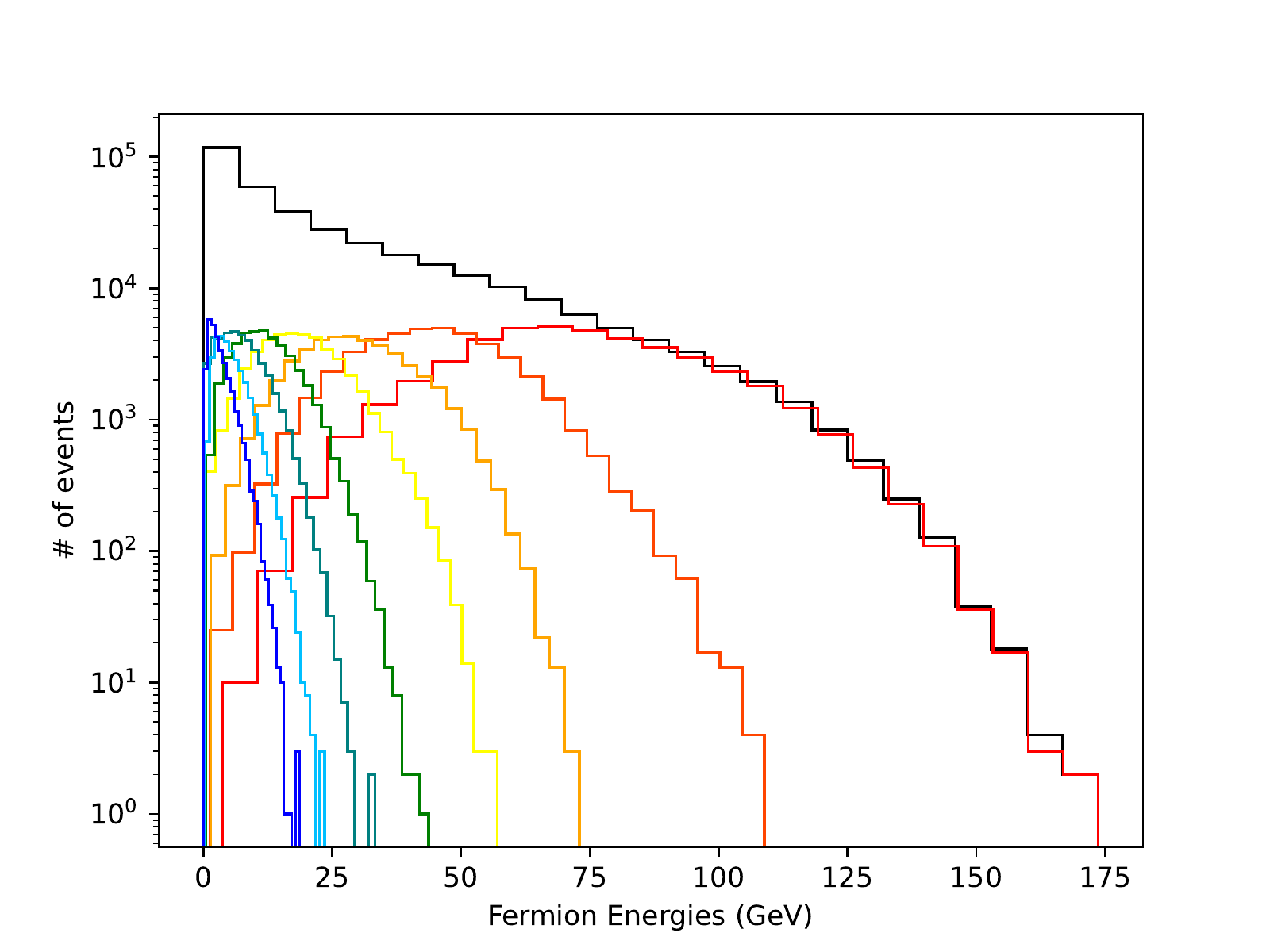}
\caption{Energy distributions of SM fermions (leptons or jets) in DM signal events. Black line is the overall fermion energy distribution. 
The colored lines show the spectra of the most energetic fermion in each event (red), the second most-energetic (orange), and so on.}
\label{fig:FermionEnergies}
\end{figure}

The energy spectra of the SM fermions in signal events are shown in Fig.~\ref{fig:FermionEnergies}. For each signal event, we order the SM fermions (including both leptons and jets) according to their energies, and show the distributions for each ranking, starting with the most energetic fermion and going down to number eight. The fermion spectra clearly reflect their origin in a cascade decay. At each step of the cascade, the decaying DM particle mass is lower than at the preceding step, leaving less energy to be transferred to fermions in the decay. As a result, the first step in each cascade typically produces the most energetic fermions, with fermion energies steadily decreasing in each subsequent step. The characteristic fermion energy pattern provides a clear observable signature for the cascade-decay origin of the signal. 

\begin{figure}
\centering
\includegraphics[width=2.8in]{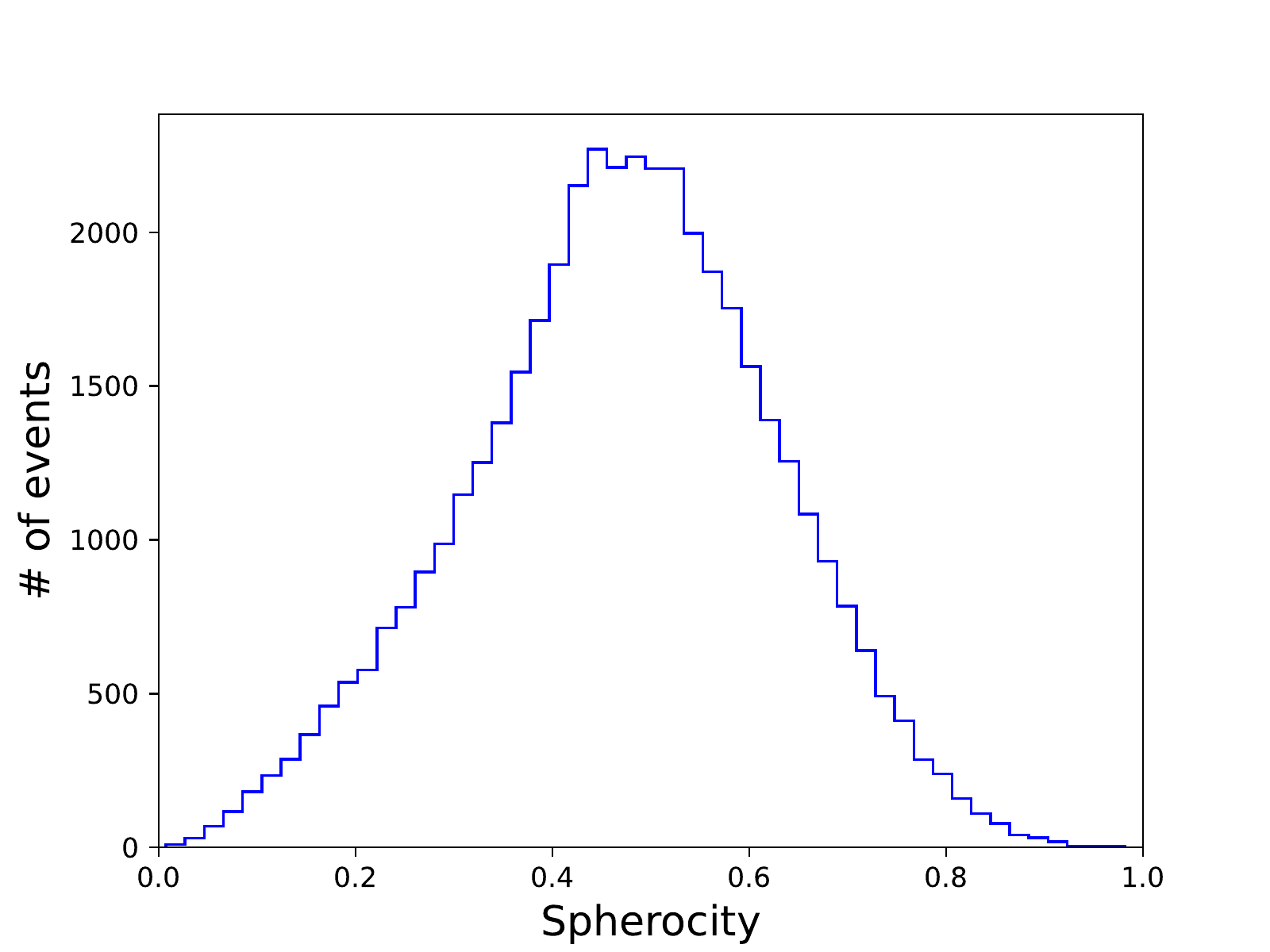}
\includegraphics[width=2.8in]{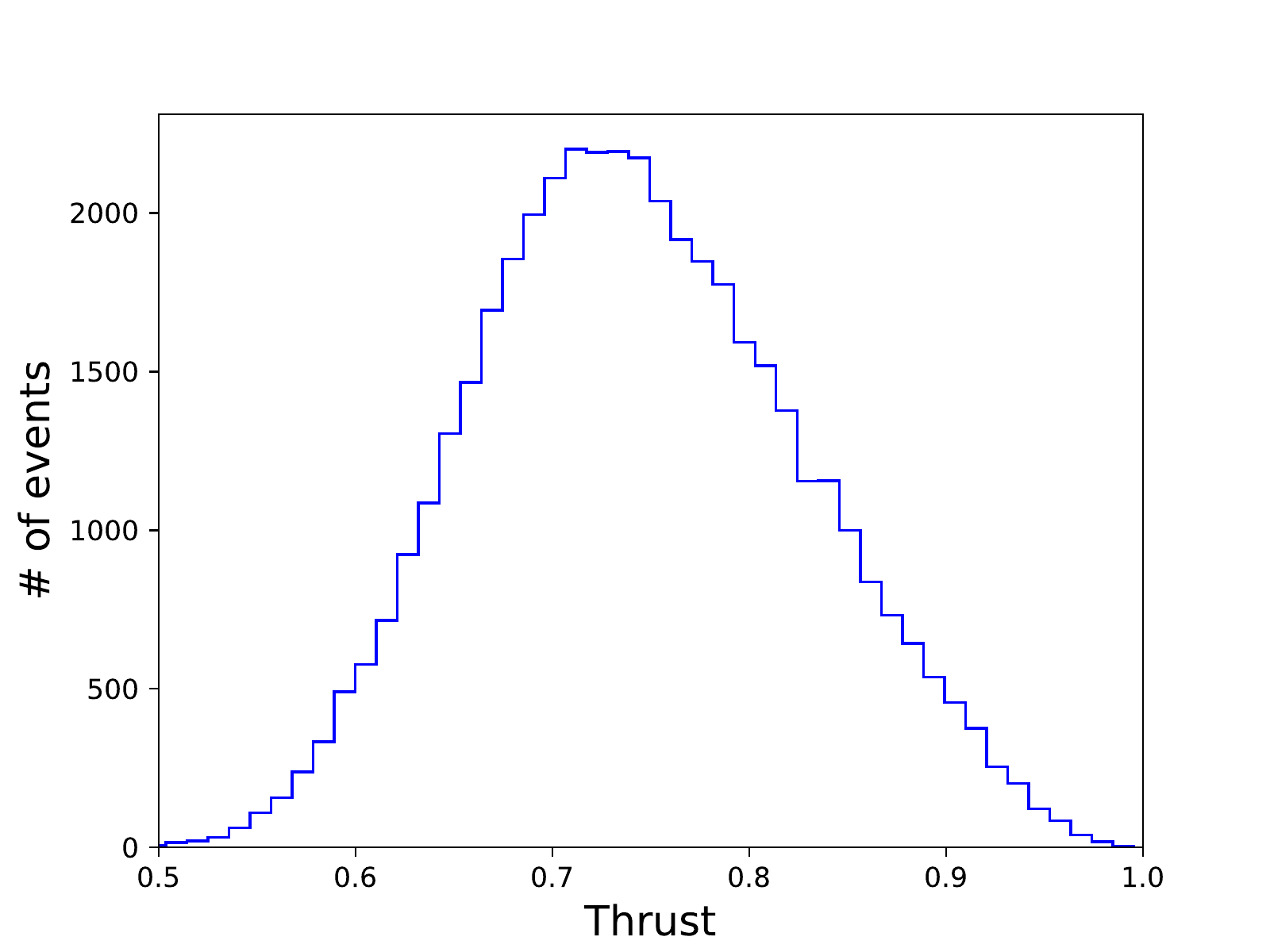}
\caption{\label{fig:epsart} Spherocity (left) and thrust (right) distributions of the signal events.}
\label{fig:Spherocity}
\end{figure}

To further characterize the geometry of the signal events, we consider the event-shape observables, spherocity $S$ and thrust $T$. These are defined as~\cite{PhysRevLett.39.1587,PhysRevLett.39.1237}
\begin{align}
    S &= 
    \bigg( 
    \frac{4}{\pi}
    \bigg)^{2}
    \text{min}_{\textbf{n}}
    \bigg( 
    \frac{\sum_{i}
    |\textbf{p}_{i}\times\textbf{n}|}
    {\sum_{i}|\textbf{p}_{i}|}
    \bigg)^{2}
    \\
    T &= \text{max}_{\textbf{n}}
    \frac{\sum_{i}
    |\textbf{p}_{i}\cdot\textbf{n}|}
    {\sum_{i}|\textbf{p}_{i}|},
\end{align}
where $i$ indexes the final-state momenta $\textbf{p}_{i}$ of all (observable) particles in the event, and $\textbf{n}$ is an arbitrary unit vector. A spherocity (thrust) value close to 1 ($\frac{1}{2}$) indicates a spherically-symmetric event, whereas a spherocity (thrust) value close to 0 (1) indicates a jet-like event. Spherocity and thrust distributions of DM signal events are shown in Figure~ \ref{fig:Spherocity}. The signal events are neither completely symmetric, nor strongly jet-like. This makes sense physically. The direction of the originally produced DM pair acts as a preferred axis for each event, but since the produced DM states are only mildly relativistic, their decay products are not strongly beamed, resulting in a rather broad distribution of momenta around this axis. 

\begin{figure}
\centering
\includegraphics[width=2.8in]{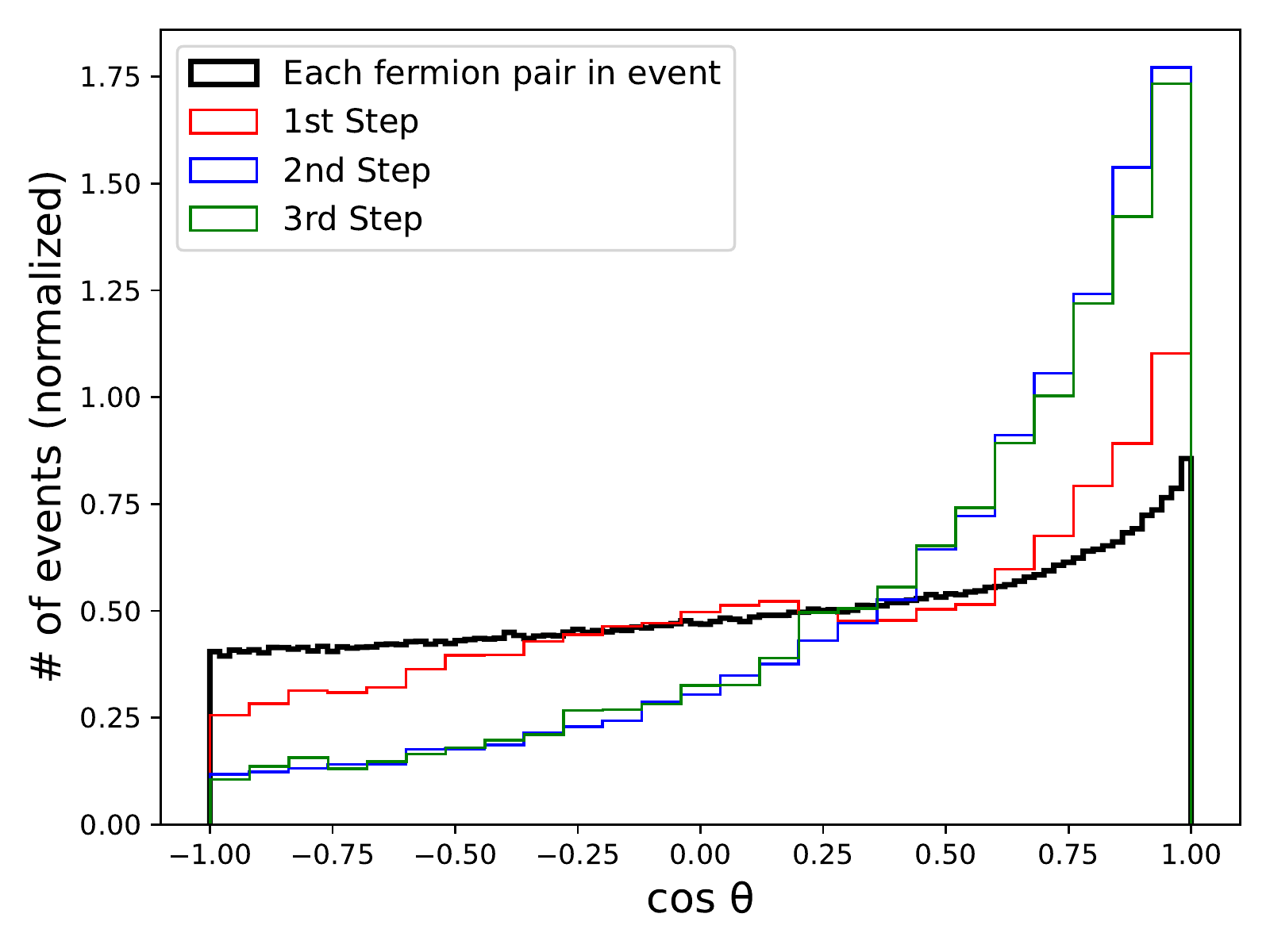}
\includegraphics[width=2.8in]{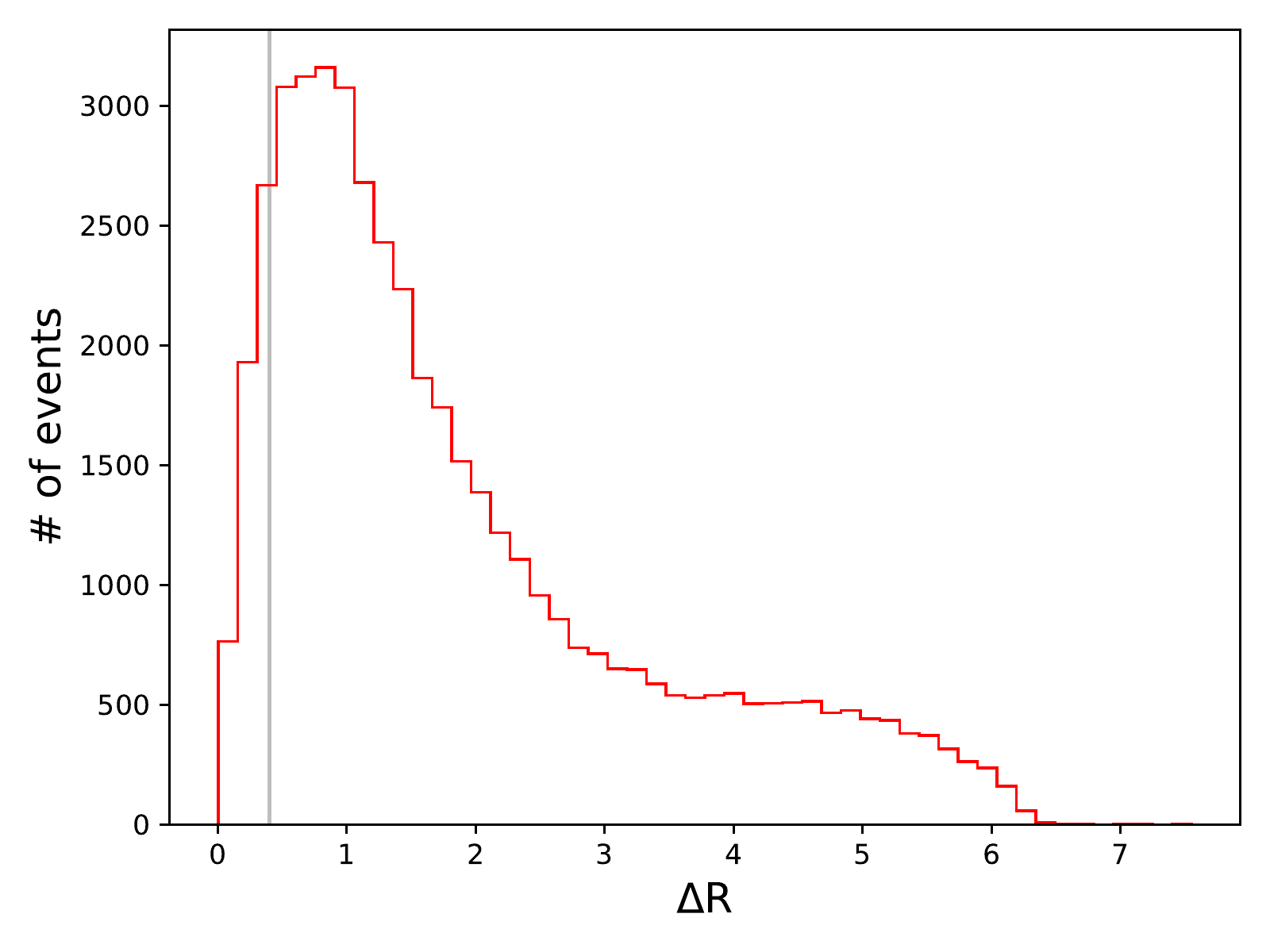}
\caption{Left panel: Angle between pairs of fermions. Black histogram includes all fermion pairs in a given event, while red/blue/green histograms correspond to fermion pairs produced in the same DM decay, in the first/second/third step in the cascade, respectively. Right panel: angular distance $\Delta R$ between quarks produced in the same DM decay (all steps in the cascade are included).}
\label{fig:FermionAngles}
\end{figure}

Another way to characterize the angular distribution of the leptons and jets produced in the cascade is by the relative angles between fermion pairs. These distributions are shown in Fig.~\ref{fig:FermionAngles}. While the overall distribution including all fermion pairs (black histogram in the left panel) is rather flat, pairs of fermions originating from the same step in the cascade tend to be approximately collinear with each other (colored histograms in the left panel). This tendency becomes more pronounced for fermions produced in later steps in a cascade decay. The decaying DM particles in these steps have higher velocities and their decay products are more boosted in the direction of the decaying particle, and hence more collimated. Given this angular correlation, it is natural to ask whether the two quarks produced in the same decay would be typically reconstructed as two separate jets, or be merged into a single jet by the jet reconstruction algorithm. This is addressed by the right panel of Fig.~\ref{fig:FermionAngles}, which shows the distribution of the quark pairs produced in the same decay as a function of $\Delta R=\sqrt{(\Delta\eta)^{2} + (\Delta\phi)^{2}}$. Most quark pairs are separated by $\Delta R>0.4$, a typical jet size used by jet reconstruction algorithms, and so would be reconstructed as two separate jets. This justifies thinking of parton-level quark multiplicity and jet multiplicity as essentially the same in these events.   

Decay widths of the near-continuum DM states span many orders of magnitude. The states with masses well above the gap scale decay promptly, while those near the gap are effectively stable on the detector time scale. A typical near-continuum DM model also contains states with decay widths in the intermediate regime, which travel macroscopic distances but decay within the detector. For example, our benchmark model contains 3 states with proper decay lengths between 1 mm and 5 m. When such states appear in the cascade, they will produce displaced vertices, giving another distinctive signature of this type of models. Since states with macroscopic lifetimes lie rather close to the gap scale, the energy of the fermion pairs associated with such displaced vertices is rather low, in the $1-10$ GeV range, and is inversely correlated with the displacement.     

\section{Conclusions and Outlook}
\label{sec:conc}

Continuum Dark Matter framework provides an alternative to traditional particle dark matter with distinct phenomenological signatures. The five-dimensional soft-wall geometry that underlies these models contains a singularity in the space-time, a finite distance away from the 4D brane. In this paper, we considered a simple toy model of how this singularity might be resolved in a fundamental theory of gravity. To this end, we introduced an IR-regulator brane, which cuts off the singular region of space-time. With this regulator, the DM spectrum becomes discrete, but mass splitting is small compared to all other physical scales in the model - a near-continuum. We then discussed the collider phenomenology of near-continuum DM. A custom-made MC tool was constructed to model production and cascade decays of DM states. Events in which DM is produced are characterized by missing energy, high multiplicity of both jets and leptons, and displaced vertices. All of these features provide a distinctive signature of this model at colliders. 

As a specific example to illustrate near-continuum DM phenomenology and validate the MC tool, we considered electron-positron collisions at $\sqrt{s}=500$~GeV. An obvious extension of this work is to generalize it to the case of hadron colliders such as the LHC. This study is currently in progress. Near-continuum DM signatures at the LHC are broadly the same as at an $e^+e^-$ collider, with some differences appearing at a quantitative level due to composite nature of initial-state particles, higher trigger thresholds for jets and leptons, etc. The distinctive nature of the near-continuum DM signal at the LHC should enable sensitive searches for this model. 

Another interesting extension of this work is to consider what happens as the IR-regulator brane is moved towards the singularity. The splitting among the dark matter KK states tends to zero in this limit, approaching the continuum spectrum. As we remarked in Section~\ref{subsubsec:GravDecay}, the width of DM decays via the $Z$-portal goes to zero in this limit, but the width of gravitational decays remains finite. This means that the narrow-width approximation eventually fails, and the KK modes can no longer be considered as asymptotic states in perturbative calculations. Methods similar to Georgi's calculations for unparticles~\cite{Georgi:2007ek,Grinstein:2008qk} can be employed to calculate inclusive DM production rates in this regime. It seems likely that evolution of the produced DM will occur entirely within the dark/gravitational sector, so that the signature of DM will be pure missing energy. An initial-state radiation photon (or gluon) would be required to render this signature visible, as is the case for direct production of ordinary particle DM at colliders~\cite{Birkedal:2004xn}. This will be considered in future work.

%%%%%%%%%%%%%%%%%%%%%%%%%%
\section*{Acknowledgments}
We would like to thank Csaba Csaki, Sungwoo Hong, Gowri Kurup and Wei Xue for collaboration and useful discussions on Continuum Dark Matter models. We are also grateful to Ameen Ismail for helpful discussions.
S.F. and M.P. are supported by the NSF grant PHY-2014071. 
S.L is supported by the Samsung Science \& Technology Foundation 
under Project Number SSTF-BA2201-06.

\appendix
\label{appendix}

\section{Simulation Framework}
\label{app:MCFramework}

In this Appendix, we describe the custom-made Monte Carlo tool used to simulate the near-continuum DM production and cascade decays. The code is available at \url{https://github.com/sferrante/VegasMC_WIC}. In addition to a standard LHE file, the output of the code is a list of event records, where each event record consist of a list of vectors of the form 
\beq
\{ \text{label}, \,\, 
   \text{PDGID}, \,\, 
    m, \,\, E, \,\, 
    p_{x}, \,\,
    p_{y}, \,\, 
    p_{z}, \,\, 
    s, \,\, \tau \}
\eeq{event_format}
where `label' specifies the step of the cascade the particle is produced at, along with the ``side" of the event ({\it i.e.} which of the two originally produced DM particles initiated the cascade). The PDGID is in the standard format for all SM particles, and PDGID=1000000 for the DM scalar. The next 5 arguments are the mass and four-momentum of the particle in GeV, while $s=0(1)$ for a stable (unstable) particle, and $\tau$ is the proper lifetime of the particle in seconds. Note that the event record includes intermediate ({\it i.e.} decayed) DM particles in the cascade, even though they are not observable. The simulated events are then used to generate plots shown in Sec.~\ref{sec:VisiblePheno}.

\subsection{Production}

The procedure for event generation starts by drawing a point $(\mu, \mu', \cos\theta)$ according to the differential production cross section discussed in Sec.~\ref{subsec:NCProduction}. This is done using {\tt VEGAS}~\cite{Lepage:1977sw}. Together with the uniformly distributed azimuthal angle $\phi$, these three variables fully determine the kinematics of the event. The four-momenta and masses of the two directly produced DM states are the first addition to the event record. 

In both production and decay, the accuracy of the simulation is improved by applying an unweighting procedure to {\tt VEGAS} samples, which is effect improves the agreement of the simulated samples with the target distributions at the expense of discarding some of the simulated events. Our unweighting algorithm is described in Sec.~\ref{app:unweight} below.

\subsection{Decay}

The produced DM states undergo cascade decay. In our simulation, we only include the decay $\phi_k \to \phi_l + Z^{(*)} \to \phi_l + f\bar{f}$, which is the dominant channel at the benchmark point used in our phenomenological study. The cascade is simulated using an iterative procedure. The 4-momentum and mass of the DM state directly produced in the $e^+e^-$ collision define the initial state for the first step of the cascade. Subsequently, the 4-momentum and mass of the DM particle in the final state of the $N$-th step in the cascade serve as the initial state for the decay at the $(N+1)$-st step. The farther into a decay chain a DM state is, the longer its lifetime will be, and eventually the state will be effectively stable with respect to the detector size. At this point, the simulation terminates and the detector-stable DM state is treated as missing energy. The condition determining this cutoff is 
\beq
\frac{\gamma v}{\Gamma(\mu)} \sim~{\rm detector~size}.  
\eeq{cascade_end}
where $v$ and $\gamma$ are the velocity and the gamma-factor of the DM particle, and $\mu$ is its mass. Assuming the detector size of a few m, we find that DM states with mass below $\mu_{\rm min}=106$~GeV can be considered detector-stable.\footnote{DM states in our simulation are not highly relativistic, see Fig.~\ref{fig:masses}, 
while dependence of the width of the DM state on its mass is very strong, so the variation of the cutoff with the DM velocity can be ignored.} We use this value to terminate the cascades in our simulation.

Each step in the cascade decay is simulated as follows. First, the code checks if the mass of the initial state particle is above $\mu_{\rm min}$; if not, the simulation is terminated. If the condition is satisfied, {\tt VEGAS} is used to draw a point $(\mu_l, x_{f}, x_{\bar{f}})$ according to the differential decay distribution discussed in Sec.~\ref{subsubsec:ZDecay}. These three variables, along with uniformly distributed angles $\theta$ and $\phi$ that fix the direction of the daughter DM scalar, determine the 4-momenta of the DM state and two fermions in the final state, in the rest frame of the decaying DM particle. These momenta are then boosted to the lab frame, using the velocity of the decaying DM particle as the boost parameter. Finally, fermion flavors are assigned according to the relevant branching fractions, fully determined by the SM $Z$ couplings.    

The differential decay rate $\frac{d^{3}\Gamma}{d\mu_{l}dx_{f}dx_{\bar{f}}}$ depends on the mass of the decaying particle $\mu_k$, which is treated as a continuum parameter in our simulation. Thus a literal implementation of the above procedure would require us to train {\tt VEGAS} separately for every decay in the event list, and each training would be used to only draw a single point. This is prohibitively computationally expensive. Since the variation of the differential decay rate with $\mu_k$ is rather slow, we can obtain a good approximation by sampling  $\frac{d^{3}\Gamma}{d\mu_l dx_{f}dx_{\bar{f}}}$ for fixed $\mu_k$ values in increments of 10 GeV within the relevant range ($\mu_k\in [100,400]$~GeV in our case). For each step in the cascade, we then draw a point from the sample with $\mu_k$ nearest to the decaying particle mass. This ``discretization" procedure generally produces a good approximation to the true distributions of the final-state particles, but  unphysical features ({\it e.g.} sharp edges) appear in some distributions due to discretization. To further improve the simulation, we perform a rescaling of the 
final-state 4-momenta and the final-state DM mass, which in effect interpolates the sampled distributions to provide a better approximation for the true decaying particle mass. The SM mass $\mu_l$ is replaced with a rescaled value $\mu_l^*$, chosen according to
\begin{align}
    \frac{\mu^{*}_{l}-\mu_{0}}
    {\mu_l-\mu_{0}}
    = 
    \frac{\mu^{\rm true}_{k}-\mu_{0}}
    {\mu^{\rm sim}_{k}-\mu_{0}}\,,
\end{align}
where $\mu^{\rm true}_{k}$ and $\mu^{\rm sim}_{k}$ refer to the actual value of the decaying DM particle and the value used in the nearest simulated sample, respectively. The fermion 4-momenta are then rescaled according to
\begin{align}
    p^\mu_{f} &\rightarrow a\,p^\mu_{f} 
    \\ 
    p^\mu_{\bar{f}} &\rightarrow \frac{1}{a}\,p^\mu_{\bar{f}}, 
\end{align} 
while the components of the final-state DM 3-momentum are rescaled according to
\beq
p^i_{\rm DM} \rightarrow b_{i}\,p^i_{\rm DM}.
\eeq{DM_mom}
The four scale factors $a$, $b_i$ are chosen so that the rescaled four-momenta satisfy energy-momentum conservation, with the actual decaying mass particle $\mu^{\rm true}_k$ as the initial energy. After rescaling, unphysical effects of discretization in distributions disappear.     

\subsection{Unweighting Procedure}
\label{app:unweight}

\begin{figure}
\centering
\includegraphics[width=3.5in]{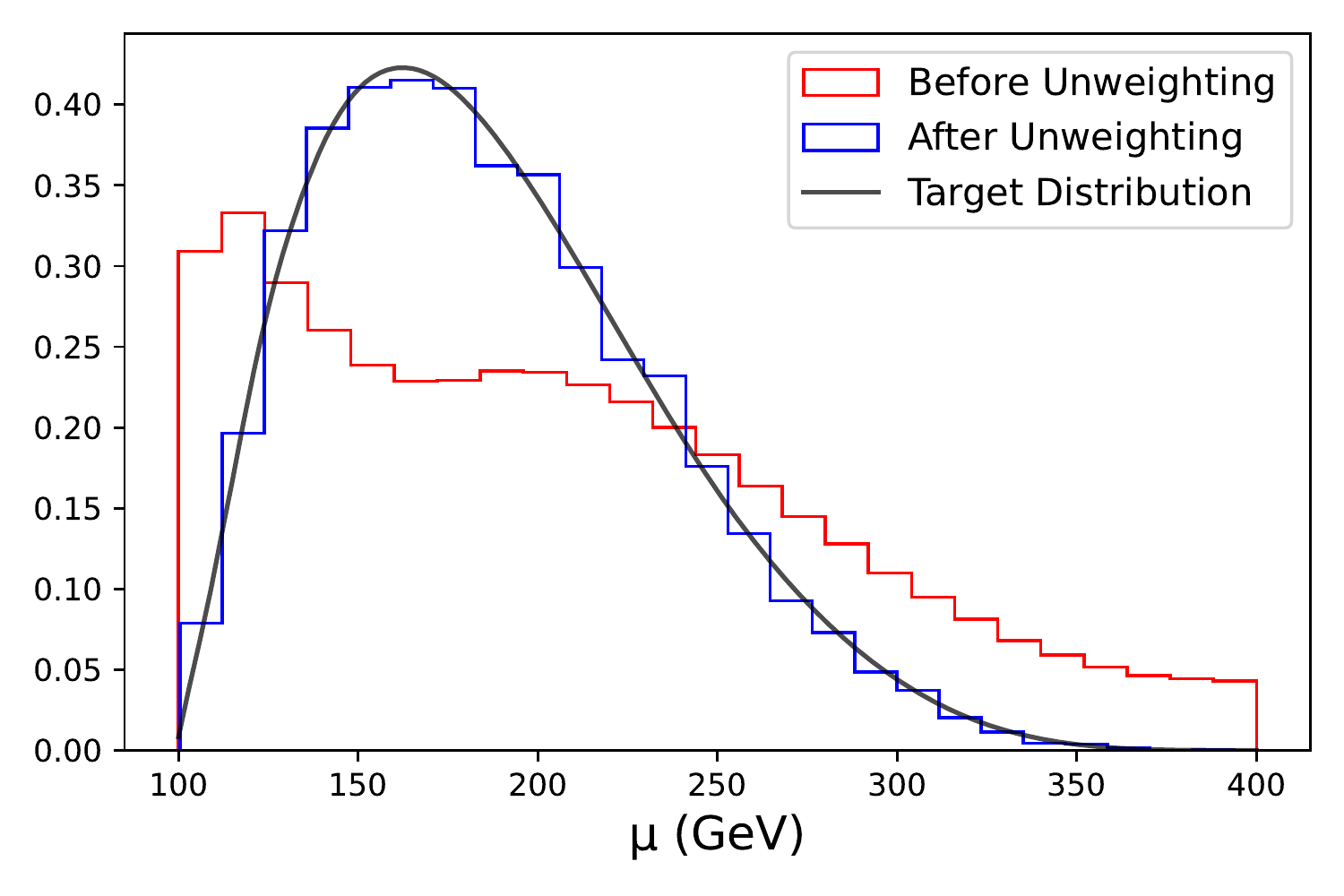}
\caption{Unweighting procedure for the simulation of the near-continuum DM production cross section as a function of the DM mass. The histograms are normalized using the integral of the target distribution, which is the mass dependent factor of Eq.~\leqn{diff_xsec}.}
\label{fig:UnweightingToy}
\end{figure}

Consider a sample of points $\{ x_i, i=1,\ldots,N\}$, where each point $x_i$ is a $d$-dimensional vector, produced by the {\tt VEGAS} algorithm trained on an (analytically known) target distribution $f(x)$. The probability distribution encoded by the sample is given by 
\begin{align}
    p(x_i) = n_{i} / N\,,
\end{align}
where $n_{i}$ is the number of points in the $d$-dimensional hypercube that contains $x_{i}$. (The hypercubes are constructed by the {\tt VEGAS} algorithm during training.) The raw weight for each point is defined by 
\begin{align}
    w_{r}(x_i) = \frac{f(x_i)}{p(x_i)}.
\end{align}
If the simulation is ideal, $w_r=1$ for all points. In practice, however, this is rarely the case. The unweighting procedure consists of removing some of the points in the simulated sample, so that the distribution of the remaining points is as close to the target $f(x_i)$ as possible. To this end, rescaled weights are defined by 
\begin{align}
    w(x_{i}) = 
    \frac{w_{r}(x_{i})}
    {\max_{i}[w_{r}(x_{i})]}\,.
\end{align}
An array of $N$ random numbers uniformly distributed between 0 to 1, denoted $r_{i}$, is then generated. The {\tt VEGAS} sample is then modified according to the rule: 
\begin{align}
    w(x_{i}) > r_{i}
    \,\,\, &\implies \,\,\, 
    \text{keep the point}
    \nonumber\\
    w(x_{i}) < r_{i}
    \,\,\, &\implies \,\,\,
    \text{discard the point}.
    \nonumber
\end{align}
The unweighting efficiency is defined by
\begin{align}
    \epsilon = N^{*} / N\,,
\end{align}
where $N^{*}$ is the number of points after unweighting. Clearly, the unweighting efficiency is higher if the original {\tt VEGAS} sample is a better approximation of the target. In our work, we find values of $\epsilon$ in the range from $\sim 20$\% for one-dimensional simulations, to $\sim 0.5$\% for sampling the three-dimensional phase space of DM decays. As an example, the result of the unweighting procedure for the DM production cross section is shown in Fig.~\ref{fig:UnweightingToy}.

\bibliographystyle{utphys}
\bibliography{ref}

\end{document}